%% file: isit11.tex
\documentclass[twocolumn,english,10pt,final,conference,letterpaper]{IEEEtran}
\usepackage[T1]{fontenc}
\usepackage{amsmath,amssymb,bbm,stmaryrd,soul}
\usepackage{tikz}
\usepackage{stfloats}
\usetikzlibrary{decorations.pathreplacing}
\usepackage{pgfplots}
\usepackage{babel}

\newcommand{\entropy}[1]{H\!\left(#1\right)}
\newcommand{\x}{\mathsf{x}}

\newcommand{\h}{\mathsf{h}}
\newcommand{\G}{\mathsf{G}}
\renewcommand{\a}{\mathsf{a}}
\renewcommand{\b}{\mathsf{b}}

\sethlcolor{red}

\makeatletter
\@addtoreset{remark}{theorem}
\makeatother

\makeatletter
\@addtoreset{corollary}{theorem}
\makeatother

\pgfplotsset{tick label style={
font=\Large}}

\begin{document}
\pgfdeclarelayer{background}
\pgfdeclarelayer{foreground}
\pgfsetlayers{background,main,foreground}
\IEEEoverridecommandlockouts
\title{Universality for the Noisy Slepian-Wolf Problem\\ via Spatial Coupling}

\author{Arvind Yedla, Henry D. Pfister, and Krishna R. Narayanan%
\thanks{This material is based upon work supported, in part, by the National Science Foundation (NSF) under Grant No. 0747470, by the Texas Norman Hackerman Advanced Research Program under Grant No. 000512-0168-2007, and by Qatar National Research Foundation.  Any opinions, findings, conclusions, and recommendations expressed in this material are those of the authors and do not necessarily reflect the views of these sponsors.}\\
  Department of Electrical and Computer Engineering\\
  Texas A\&M University\\
  Email: \{yarvind,hpfister,krn\}@tamu.edu}
\maketitle
\begin{abstract}
  We consider a noisy Slepian-Wolf problem where two correlated sources
  are separately encoded and transmitted over two independent binary
  memoryless symmetric channels. Each channel capacity is assumed to be
  characterized by a single parameter which is not known at the
  transmitter. The receiver has knowledge of both the source correlation
  and the channel parameters. We call a system universal if it retains
  near-capacity performance without channel knowledge at the
  transmitter.

  Kudekar et al.~recently showed that terminated low-density
  parity-check (LDPC) convolutional codes (a.k.a. spatially-coupled LDPC
  ensembles) can have belief-propagation thresholds that approach their
  maximum a-posteriori thresholds.  This was proven for binary erasure
  channels and shown empirically for binary memoryless symmetric
  channels. They also conjectured that the principle of spatial coupling
  is very general and the phenomenon of threshold saturation applies to
  a very broad class of graphical models.  In this work, we derive an
  area theorem for the joint decoder and empirically show that threshold
  saturation occurs for this problem. As a result, we demonstrate
  near-universal performance for this problem using the proposed
  spatially-coupled coding system. A similar result is also discussed
  briefly for the 2-user multiple-access channel.
\end{abstract}

\begin{IEEEkeywords}
  LDPC codes, spatial coupling, EXIT functions, density evolution,
  correlated sources, non-systematic encoders, joint decoding,
  protograph, area theorem.
\end{IEEEkeywords}
\thispagestyle{empty}

\pagestyle{empty}

\section{Introduction}
\label{sec:intr-backgr}
\input{intro}

\section{Density Evolution and (G)EXIT Curves}
\label{sec:dens-evol-exit}
\input{de_exit}

\section{Spatial Coupling}
\label{sec:spatial-coupling}
\input{spatial_coupling}

\begin{figure*}[t!]
  \centering
  \begin{minipage}[t]{0.32\linewidth}
      \centering
  \input{acpr_36_spatial_mac}
  \vspace{-3mm}
  \caption{DE ACPR of a spatially coupled
     $(3,6,32,4)$ LDPC code for the two user
     Gaussian MAC. The DE
    results for the regular LDPC$(3,6)$ ensemble are shown for
    comparison.}
  \label{fig:roc_ldpc_awgn_bac}
  \end{minipage}
  \begin{minipage}[t]{0.32\linewidth}
  \centering
  \input{46_lrlw_acpr}
  \vspace{-3mm}
  \caption{DE ACPR of the spatially
    coupled punctured $(4,6,64,10)$ LDPC and the regular punctured
    LDPC$(4,6)$ ensembles for transmission over
    erasure channels with erasure correlated sources.}
  \label{fig:roc_ldpc_erasure}
  \end{minipage}
  \begin{minipage}[t]{0.32\linewidth}
  \centering
  \input{46_lrlw_acpr_awgn}
  \vspace{-3mm}
  \caption{DE ACPR of the spatially
    coupled punctured $(4,6,64,10)$ LDPC and the regular punctured
    LDPC$(4,6)$ ensembles for transmission over AWGN
    channels with BSC correlated sources.}
  \label{fig:roc_ldpc_awgn}
  \end{minipage}
\end{figure*}

\section{The $2$-user Gaussian MAC}
\label{sec:2-user-gaussian}
\input{gaussian_mac}

\section{Results and Concluding Remarks}
\label{sec:results}
\input{results}

\bibliographystyle{IEEEtran}

\end{document}

%% file: intro.tex
The phenomenon of threshold saturation via spatial coupling was
introduced in \cite{Kudekar-it11,Kudekar-istc10} to
describe the excellent performance of convolutional LDPC codes over
binary-input memoryless symmetric (BMS) channels
\cite{Lentmaier-it05}. Kudekar et al.~prove that the belief
propagation (BP) threshold of the spatially coupled ensemble is
\emph{essentially} equal to the maximum a-posteriori (MAP) threshold
of the underlying ensemble when transmission takes place over a binary
erasure channel (BEC) \cite{Kudekar-it11}. Empirical evidence
of this phenomenon for BMS channels has been observed in
\cite{Kudekar-istc10,Lentmaier-it05}.

The underlying principle behind the impressive performance of
spatially-coupled codes is very broad and Kudekar~et~al. conjecture
that the same phenomenon occurs for more general channels. In this
work, we consider a noisy Slepian-Wolf problem. The outputs of two
discrete memoryless correlated sources, $\left(U_1,U_2\right)$, are
transmitted to a central receiver through two independent discrete
memoryless channels with capacities $C_1$ and $C_2$, respectively. In
\cite{Yedla-aller09}, the authors consider the noisy Slepian-Wolf problem
and observed that the MAP threshold of the punctured LDPC$(4,6)$
ensemble is very close to the capacity region for transmission over
erasure channels.
Therefore, the phenomenon of threshold saturation motivates the use of
spatial coupling as a potentially universal coding scheme for the noisy
Slepian-Wolf problem. In this paper, this observation is extended to
the $2$-user Gaussian multiple access channel (MAC).

We will assume that the channels belong to the same channel family, and
that each channel can be parameterized by a single parameter $\alpha$
(e.g., the erasure probability for erasure channels). We also assume
that the channel parameters are not known at the transmitter. The system
model is shown in Fig.~\ref{fig:sys-model}. The two encoders are not
allowed to communicate and hence they must use independent encoding
functions. We also assume that both the encoders use identical rates
$R=k/n$, i.e., they map $k$ input symbols $(\mathbf{U}_1\text{ and }
\mathbf{U}_2)$ to $n$ output symbols $(\mathbf{X}_1\text{ and }
\mathbf{X}_2)$, respectively. The decoder receives
$(\mathbf{Y}_1,\mathbf{Y}_2)$ and computes an estimate of
$(\mathbf{U}_1,\mathbf{U}_2)$. 

\begin{figure}[!h]
  \centering
  \input{system_model.tex}
  \caption{System Model}
  \label{fig:sys-model}
\end{figure}

Reliable transmission over a channel pair $(\alpha_1,\alpha_2)$ is
possible as long as the pair satisfies the Slepian-Wolf conditions 

\begin{equation}
  \label{eq:sw}
  \begin{split}
    \frac{C_1 (\alpha_1)}{R} &\geq \entropy{U_1\middle |U_2}, \\
    \frac{C_2 (\alpha_2)}{R} &\geq \entropy{U_2\middle |U_1}\text{ and} \\
    \frac{C_1 (\alpha_1)}{R}+\frac{C_2 (\alpha_2)}{R}&\geq
    \entropy{U_1,U_2}
  \end{split}
\end{equation}
are satisfied. For a given pair of rate-$R$ encoding functions and a
joint decoding algorithm, we say that a pair of channel parameters
$(\alpha_1,\alpha_2)$ is \emph{achievable} if the encoder/decoder
combination can achieve an arbitrarily low probability of error as the
blocklength $n \rightarrow \infty$. As in \cite{Yedla-istc10}, the
achievable channel parameter region (ACPR) is defined as the set of
all channel parameters which are achievable, and the Slepian-Wolf
region (illustrated in Fig.~\ref{fig:sw} for erasure channels) is the
set of all channel parameters $(\alpha_1,\alpha_2)$ for which
(\ref{eq:sw}) is satisfied. Coding schemes for which the ACPR is equal
to the Slepian-Wolf region are said to be \emph{universal}. Such
schemes are important because, in some practical situations, it is
unreasonable to have knowledge of the channel parameters at the
transmitter. Hence a single coding scheme needs to perform well over a
large set of channel parameters.

\begin{figure}[!htb]
  \centering
  \input{slepian_wolf_region.tex}
  \caption{The Slepian-Wolf region for erasure channels for a
    rate pair $(R,R)$.}
  \label{fig:sw}
\end{figure}

In this paper, we consider the following scenarios:
\begin{enumerate}
\item The channels are binary erasure channels (BECs) and the source
  correlation is modeled through erasures. Let $Z$ be a Bernoulli-$p$
  random variable. The sources $U_1$ and $U_2$ are defined by
\begin{equation*}
  \left(U_1,U_2\right) = \left\{ \begin{array}{l}
      \textnormal{i.i.d. Bernoulli $\frac{1}{2}$ r.v.s}, \textnormal{if } Z=0\\\\
      \textnormal{same Bernoulli $\frac{1}{2}$ r.v. $U$ },
      \textnormal{if } Z=1
    \end{array} \right.
\end{equation*}
This gives $\entropy{U_1|U_2} = \entropy{U_2|U_1} = 1-p$ and
$\entropy{U_1,U_2} = 2-p$. In this model, the decoder has access to
the side information $Z$. While this model is not realistic, it is
useful as a toy model that enables us to gain a better understanding
of the problem.
\item A more realistic model is one where the channels are
  binary-input additive white Gaussian noise channels (BAWGNC) and the
  source correlation is modeled through a virtual correlation channel
  analogous to a binary symmetric channel (BSC). It is useful to
  visualize this correlation by the presence of an auxiliary binary
  symmetric channel (BSC) with parameter $1-p$ between the sources. In
  other words, $U_2$ is the output of a BSC with input $U_1$ (a
  Bernoulli-$1/2$ random variable) i.e., $U_2 = U_1 + Z$. Here $Z$ is
  a Bernoulli-($1-p$) random variable and can be thought of as an {\em
    error}. Let $H_2(\cdot)$ denote the binary entropy function. Then,
  $\entropy{U_1|U_2} = \entropy{U_2|U_1} = H_2(p)$ and
  $\entropy{U_1,U_2} = 1 + H_2(p)$. In this model, the side
  information $Z$ is not available at the decoder.
\end{enumerate}

Although separation between source and channel coding is known to be
optimal for this problem \cite{Barros-it06}, it can be
beneficial to take a joint source-channel coding approach (via direct
channel coding and joint decoding at the receiver)
\cite{GarciaFrias-dcc01}. This problem is considered in
\cite{Abrardo-ita09,Martalo-ita10}, where the authors choose a
code that performs well at one point on the Slepian-Wolf region and
evaluate its performance for different channel parameters. As a
result, the performance of the code is far from the optimal
performance for some channel parameters. Even the optimized degree
profiles of LDPC codes for this problem are far from universal
\cite{Yedla-istc10}. In this paper, we derive the area theorem for
the joint decoder and compute (G)EXIT curves for transmission over
symmetric channel conditions. This provides empirical evidence that
the phenomenon of threshold saturation occurs for the noisy
Slepian-Wolf problem. Moreover, density evolution (DE) results suggest
that the spatially-coupled punctured-systematic LDPC$(4,6)$ ensemble
is near universal.

%% file: system_model.tex
\begin{tikzpicture}[scale=0.42,>=stealth,xshift=-1cm]
\draw (0,0) rectangle +(3,1);
\draw (1.5,.5) node {\footnotesize Source $2$};
\draw (0,4) rectangle +(3,1);
\draw (1.5,4.5) node {\footnotesize Source $1$};
\draw (5,0) rectangle +(3,1);
\draw (6.5,.5) node {\footnotesize Encoder $2$};
\draw (6.5,1.5) node {\footnotesize $R$};
\draw (5,4) rectangle +(3,1);
\draw (6.5,4.5) node {\footnotesize Encoder $1$};
\draw (6.5,5.5) node {\footnotesize $R$};
\draw (10,0) rectangle +(3,1);
\draw (11.5,.5) node {\footnotesize Channel $2$};
\draw (11.5,1.5) node {\footnotesize $C_2$};
\draw (10,4) rectangle +(3,1);
\draw (11.5,4.5) node {\footnotesize Channel $1$};
\draw (11.5,5.5) node {\footnotesize $C_1$};
\draw (15,2) rectangle +(3,1);
\draw (16.5,2.5) node {\footnotesize Decoder};

\draw[->] (3,0.5) -- (5,0.5) node[pos=0.6,above] {$\scriptstyle{\mathbf{U}}_2$};
\draw[->] (3,4.5) -- (5,4.5) node[pos=0.6,above] {$\scriptstyle{\mathbf{U}}_1$};
\draw[->] (8,0.5) -- (10,0.5) node[midway,above] {$\scriptstyle{\mathbf{X}}_2$};
\draw[->] (8,4.5) -- (10,4.5) node[midway,above] {$\scriptstyle{\mathbf{X}}_1$};
\draw[->] (13,0.5) -- (15,2.25) node[sloped,midway,above] {$\scriptstyle{\mathbf{Y}}_2$};
\draw[->] (13,4.5) -- (15,2.75) node[sloped,midway,above] {$\scriptstyle{\mathbf{Y}}_1$};
\draw[dashed] (-0.4,-0.4) rectangle +(3.9,5.9);
\draw (1.5,2.75) node {\footnotesize Correlated};
\draw (1.5,2.25) node {\footnotesize Sources};
\end{tikzpicture}


%% file: slepian_wolf_region.tex
\begin{tikzpicture}[>=stealth,scale=0.9]
\draw[->] (-0.5,0) -- (5,0) node[very near end,sloped,below] {$\qquad\epsilon_1$};
\draw[->] (0,-0.5) -- (0,5) node[very near end,left] {$\qquad\epsilon_2$};
\shade[top color=gray!30!white, bottom color=gray!30!white] (0,0) -- (0,3.5) -- (1.5,3.5) -- (3.5,1.5) -- (3.5,0) -- cycle;
\draw[thick] (0,0) -- (0,3.5) -- (1.5,3.5) -- (3.5,1.5) -- (3.5,0) -- cycle;
\draw[gray, very thin] (2.425,2.425) -- (2.575,2.575);
\draw[very thick] (3.5cm,-2pt) -- (3.5cm,2pt) node[below=5pt] {$\scriptstyle 1-H(U_1\mid U_2)R$};
\draw[very thick] (1.5cm,-2pt) -- (1.5cm,2pt) node[below=5pt] {$\scriptstyle{1-H(U_1)R}$};
\draw[very thick] (-2pt,3.5cm) -- (2pt,3.5cm) node[left=5pt] {$\scriptstyle 1-H(U_2\mid U_1)R$};
\draw[very thick] (-2pt,1.5cm) -- (2pt,1.5cm) node[left=5pt] {$\scriptstyle{1-H(U_2)R}$};
\draw[step=.5cm,gray,very thin] (-0.3,-0.3) grid (4.8,4.8);
\draw (3,2.5) node[above] {symmetric channel condition};
\filldraw[gray] (2.5,2.5)
circle (2pt);
\end{tikzpicture}


%% file: de_exit.tex
We assume that the sequences $\mathbf{U}_1$ and $\mathbf{U}_2$ are
encoded using a punctured-systematic encoder for LDPC codes whose
degree distribution functions are given by
$\left(\lambda,\rho\right)$. The advantages of using a
punctured-systematic encoder are discussed in
\cite{Yedla-istc10}. Based on standard notation \cite{RU-2008}, we
let $\lambda(x) = \sum_i \lambda_i x^{i-1}$ be the degree distribution
(from an edge perspective) corresponding to the variable nodes and
$\rho(x) = \sum_i \rho_i x^{i-1}$ be the degree distribution (from an
edge perspective) of the parity-check nodes in the decoding graph. The
coefficient $\lambda_i$ (resp. $\rho_i$) gives the fraction of edges
that connect to the variable nodes (resp. parity-check nodes) of
degree $i$. Likewise, $L_i$ is the fraction of variable nodes with
degree $i$. The fraction of punctured (i.e., systematic) bits is given
by
\begin{align}
\label{eq:punc_fraction}
  \gamma \triangleq 1 -
    \frac{\int_0^1\rho(x)\,\text{d}x}{\int_0^1\lambda(x)\,\text{d}x}.
\end{align}

The remainder of this section makes heavy use of the terminology and
notation from \cite{RU-2008} for DE analysis and (G)EXIT curves.  Let
$\a_{\ell}$ and $\b_{\ell}$ denote the $L$-density\footnote{Assuming
  that the transmission alphabet is $\{\pm 1\}$, the densities are
  conditioned on the transmission of a $+1$.} of the messages
emanating from the variable nodes at iteration $\ell$, corresponding
to codes $1$ and $2$. The density evolution (DE) equations
\cite{Yedla-istc10} can be written as follows
\begin{equation}
  \label{eq:de-vc}
  \begin{split}
    \a_{\ell+1} &= \left[\gamma
      f\Bigl(L\left(\rho(\b_{\ell})\right)\Bigr) +
      (1-\gamma)\a_{\text{BMSC}}\right] \varoast \lambda(\rho(\a_{\ell})) \\
    \b_{\ell+1} &= \left[\gamma
      f\Bigl(L\left(\rho(\a_{\ell})\right)\Bigr) +
      (1-\gamma)\b_{\text{BMSC}}\right] \varoast \lambda(\rho(\b_{\ell})), \\
  \end{split}
\end{equation}
where $\lambda(\a)=\sum_i\lambda_i\a^{\varoast(i-1)}$, $L(\a)=\sum_i
L_i\a^{\varoast(i-1)}$, $\rho(\a)=\sum_i\rho_i\a^{\boxast(i-1)}$,
$\a_{\text{BMSC}}$ and $\b_{\text{BMSC}}$ are the densities of the
log-likelihood ratios received from the channel. The operators
$\varoast$ and $\boxast$ are the standard density transformation
operators at the variable and check nodes respectively
\cite{RU-2008}. The operator $f$ at the correlation nodes depends on
the equivalent channel corresponding to the correlation model, as
described in \cite{Chen-isit06}.  For the correlation models
considered, one can derive a generalized symmetry condition that allows
the function $f$ to be chosen so that DE can be performed under the
all-zero codeword assumption.

In Sections~\ref{sec:erasure-correlation} and
\ref{sec:bsc-correlation}, we describe the (G)EXIT curves for the
joint decoder. For simplicity, we consider (G)EXIT curves for
symmetric channel conditions throughout this work. In this case, the
DE equations collapse into the recursion $\a_{\ell + 1} =
\mathsf{D}(\a_{\text{BMSC}},\a_{\ell})$. Similar to single
user channels, the area theorem for the joint decoder can be used to
compute an upper bound on the MAP threshold of the joint decoder.  For
example, this technique was applied to the joint decoding of a
finite-state channel and an LDPC code in \cite{Wang-turbo08}. It has
been observed that this upper bound is tight for regular LDPC
ensembles transmitted over the BEC \cite{Measson-it08}. For asymmetric
channel conditions, we can define $2$-dimensional (G)EXIT surfaces
analogously and the area theorem gives outer bounds on the MAP
boundary. As described in \cite{RU-2008}, it is useful to assume that
each bit of user $1$ (user $2$) has been transmitted through a channel
with parameter $\alpha_{i}^{(1)}$ ($\alpha_{i}^{(2)}$), and suppose
that these parameters are differentiable functions of a common
parameter $\alpha$. The area theorem follows trivially from the
definition of the MAP (G)EXIT function and is given by
\begin{align*}
  \int_{\alpha^{MAP}}^{1}h^{MAP}(\alpha)\text{d}\alpha =
  \frac{\gamma H(U_{1},U_{2})}{2(1-\gamma)}.
\end{align*}

\subsection{Erasure Correlation}
\label{sec:erasure-correlation}
For erasure correlation with probability $p$, there is a parity-check
at the correlation-node with probability $p$ and with probability
$1-p$ there is no parity-check, so $f(\a) = (1-p) + p\a$. For
simplicity, we consider the EXIT curves for the case when the channel
erasure probabilities for both users are equal. The extended belief
propagation (EBP) EXIT curve for the joint decoder with symmetric
channel conditions, is given in parametric form by
\begin{align*}
  h^{{\text{EBP}}} &= (\epsilon(x),L(1 - \rho(1 -
  x))),\,x\in[0,1],\;\text{where} 
\end{align*}
\begin{align*}
    \epsilon(x) &= \frac{1}{1-\gamma}\left[\frac{x}{\lambda(1-\rho(1-x))} - \gamma f\Bigl(L(1 - \rho(1 - x))\Bigr)\right].
\end{align*}
The EBP EXIT curve and the MAP threshold for the punctured LDPC$(4,6)$
ensemble are shown in Fig.~\ref{fig:spatial-ebp-exit-lrlw}. The MAP
threshold at symmetric channel conditions is $\epsilon^{\text{MAP}}
\approx 0.6245$, while the Slepian-Wolf bound is $\epsilon = 0.625$.

\subsection{BSC Correlation}
\label{sec:bsc-correlation}
Using the generalized symmetry condition for this model results in the
correlation node function $f(\a) = \a_{\text{BSC}(p)}\boxast\a$.  It
turns out that the GEXIT kernel for the joint decoder is the same as
that of the standard BAWGNC, given by
\begin{align*}
  l^{\a_{\text{BAWGNC}(\h)}}(y)\! &=\!
  \left(\!\int\frac{\text{e}^{{-\frac{(2-2/\sigma^{2})^{2}\sigma^{2}}{8}}}}{1+\text{e}^{z+y}}\text{d}z\!\right)\!\!/\!\!\left(\!\int\frac{\text{e}^{{-\frac{(2-2/\sigma^{2})^{2}\sigma^{2}}{8}}}}{1+\text{e}^{z}}\text{d}z\!\right),
\end{align*}
where $\sigma$ is the unique positive number such that the BAWGNC with
noise variance $\sigma^{2}$ has entropy $\h$. Let
\begin{align*}
  \G(\a_{\text{BAWGNC}(\h)},\a) = \int \a(y) l^{\a_{\text{BAWGNC}(\h)}}(y)\text{d}y
\end{align*}
be the GEXIT functional applied to the density $\a$. For each fixed
point (not necessarily stable) of density evolution
$(\a_{\text{BAWGNC}(\h)},\a)$, satisfying $\a =
\mathsf{D}(\a_{\text{BAWGNC}(\h)},\a)$, the point
$(\h,\G(\a_{\text{BAWGNC}(\h)},\a))$ lies on the EBP GEXIT curve. The
EBP GEXIT curve is the set of all these points and can be computed
numerically as outlined in \cite{Measson-it09}. The EBP
GEXIT curve and the MAP threshold for the punctured LDPC$(4,6)$
ensemble are shown in Fig.~\ref{fig:spatial-ebp-gexit-lrlw}. The MAP
threshold for symmetric channel conditions is $\h^{\text{MAP}} \approx
0.6324$ and the Slepian-Wolf bound is $\h \approx 0.6328$.


%% file: spatial_coupling.tex
Spatial coupling is best described by the $(l,r,L)$ ensemble through a
protograph \cite{Kudekar-it11,Sridharan-aller04}. We
briefly review the protograph structure at the joint decoder
here. Consider the protograph of a standard LDPC$(4,6)$
ensemble. There are two check nodes and three variable nodes. For each
user, take a collection of $(2L+1)$ protographs at positions
$[-L,L]\triangleq \{-L,\cdots,L\}$ and couple them as described in
\cite{Kudekar-it11}. One variable node at each position
$i\in[-L,L]$ from the first user is punctured and connected to a
punctured variable node at the same position of the second user. The
resulting protograph, shown in Fig.~\ref{fig:protograph}, is then
expanded $M$ times to form the parity-check matrix of the joint
system. This structure is fundamental to the phenomenon of threshold
saturation observed at the joint decoder. It is simply not sufficient
to use spatially-coupled codes with random connections between the
information nodes. Such a coupling will only result in pushing the
threshold of the component codes to the MAP threshold, but may have
little effect on the BP threshold of the joint system.

\begin{figure}[htb]
  \centering
  \input{protograph_jd}
  \caption{Protograph of the joint decoder}
  \label{fig:protograph}
\end{figure}
Although this ensemble is very instructive in understanding the
universality of spatially-coupled codes, the EBP curves for this
ensemble exhibit wiggles around the MAP threshold (similar to the single
user channels as discussed in \cite{Kudekar-it11}) 
, for the case of transmission over erasure channels. The magnitude of
these wiggles appears to remain constant with increasing $L$ and their
presence implies that the BP threshold is smaller than the MAP threshold
of the underlying ensemble. Therefore, the $(4,6,L)$ ensemble cannot be
universal. To overcome this, we use the $(l,r,L,w)$ ensemble introduced
in \cite{Kudekar-it11} for the remainder of this work.


\subsection{The $(l,r,L,w)$ ensemble}
\label{sec:l-r-l-1}
The $(l,r,L,w)$ spatially-coupled ensemble can be described as follows:
Place $M$ variable nodes at each position in $[-L,L]$. The check nodes
are placed at positions $[-L,L+w-1]$, with $\frac lr M$ check nodes at
each position. The connections are made as described in
\cite{Kudekar-it11}. This procedure
generates a Tanner graph for the $(l,r,L,w)$ ensemble.

For this work we consider codes of rate $1/3$, punctured to a rate
$1/2$. Two such graphs (generated by the above procedure) are taken
and $2M/3$ variable nodes ($M/3$ from each graph) at each position are
connected by a random (uniform) permutation of size $M/3$ via
correlation nodes. This procedure ensures that all the variable node
positions are symmetric (as opposed to Fig.~\ref{fig:protograph})
with respect to puncturing and correlation, enabling us to write down
the density evolution (DE) equations as described in the following
section.

\subsection{Density Evolution of the $(l,r,L,w)$ Ensemble}
\label{sec:density-evolution}
Let $\a_{i}^{(\ell)}$ and $\b_{i}^{(\ell)}$ denote the average density
emitted by the variable node at position $i$, at iteration $\ell$, for
codes $1$ and $2$ respectively. Let $\Delta_{+\infty}$ denote the delta
function at $+\infty$ and set $\a_{i}^{(\ell)} = \b_{i}^{(\ell)} =
\Delta_{+\infty}$ for $i\notin[-L,L]$. The channel densities for codes
$1$ and $2$ are denoted by $\a_{\text{BMSC}}$ and $\b_{\text{BMSC}}$
respectively. All the above densities are $L$-densities conditioned on
the transmission of the all-zero codeword (see
Section~\ref{sec:dens-evol-exit}). We consider the parallel schedule for
each user (as described in \cite{Kudekar-it11}) and update the
correlation nodes before proceeding to the next iteration. Let us define
\begin{align*}
  \mbox{\small $g(\x_{i-w+1},\cdots,\x_{i+w-1})\! \triangleq\! \left(\!\displaystyle\frac
    1w\!\sum_{j=0}^{w-1}\!\left(\!\frac 1w\!\displaystyle\sum_{k=0}^{w-1}\x_{i+j-k}\right)^{\boxast(r-1)}\right)^{\varoast(l-1)}\!\!,$}
\end{align*}
\begin{align*}
    \mbox{\small $\Gamma(\x_{i-w+1},\cdots,\x_{i+w-1})\! \triangleq\! \left(\!\displaystyle\frac
    1w\!\sum_{j=0}^{w-1}\!\left(\!\frac
      1w\!\displaystyle\sum_{k=0}^{w-1}\x_{i+j-k}\right)^{\boxast(r-1)}\right)^{\varoast
    l}\!\!.$}
\end{align*}
The DE equations for the joint spatially-coupled system
can be written as
\begin{align*}
  \a_{i}^{(\ell+1)} &= \bigl[\gamma
    f\left(\Gamma(\b_{i-w+1}^{(\ell)},\cdots,\b_{i+w-1}^{(\ell)})\right) +
    (1-\gamma)\a_{\text{BMSC}}\bigr]\varoast \\
    &\phantom{==[}g(\a_{i-w+1}^{(\ell)},\cdots,\a_{i+w-1}^{(\ell)}),\\
  \b_{i}^{(\ell+1)} &= \bigl[\gamma
    f\left(\Gamma(\a_{i-w+1}^{(\ell)},\cdots,\a_{i+w-1}^{(\ell)})\right) +
    (1-\gamma)\b_{\text{BMSC}}\bigr]\varoast \\
    &\phantom{==[}g(\b_{i-w+1}^{(\ell)},\cdots,\b_{i+w-1}^{(\ell)}),
\end{align*}
for $i\in[-L,L]$. For a further discussion of the DE
equations for the $(l,r,L,w)$ spatially-coupled ensembles on BMS
channels, see \cite{Kudekar-istc10}. Let $\underbar{a} =
(\a_{-L},\cdots,\a_{L})$. The fixed points of DE for
symmetric channel conditions are given by
$(\a_{\text{BMSC}(\h)},\underbar{a})$, which satisfy
\begin{align}
\label{eq:de-sym}
  \a_{i} &= \bigl[\gamma
    f\left(\Gamma(\a_{i-w+1},\cdots,\a_{i+w-1})\right) +
    (1-\gamma)\a_{\text{BMSC}(\h)}\bigr]\varoast\notag \\
    &\phantom{==[}g(\a_{i-w+1},\cdots,\a_{i+w-1}).
\end{align}
We can use the procedure outlined in \cite{Measson-it09} to
compute both the stable and unstable fixed points which satisfy
(\ref{eq:de-sym}). Define
\begin{align*}
  \G(\a_{\text{BMSC}(\h)},\underbar{a}) &= \frac{1}{2L+1}\sum_{i=-L}^{L}\G(\a_{\text{BMSC}(\h)},\a).
\end{align*}

The EBP GEXIT curve is the set of points
$(\h,\G(\a_{\text{BMSC}(\h)},\underbar{a}))$. The resulting curves for
the erasure channel with erasure correlated sources are shown in
Fig.~\ref{fig:spatial-ebp-exit-lrlw} and those for the AWGN channel
with BSC correlated sources are shown in
Fig.~\ref{fig:spatial-ebp-gexit-lrlw}. These curves are very similar
to the single user case and demonstrate the phenomenon of threshold
saturation at the joint decoder, for symmetric channel conditions. For
channel parameters not on the symmetric line, these plots imply
threshold saturation towards the MAP boundary.


\begin{figure}[t!]
  \centering
  \input{ebp_exit_46_lrlw_joint_nsw}
  \caption{EBP EXIT curves of the $(4,6,L,w)$ and
    $(4,6)$ ensembles for transmission over erasure channels with
    erasure correlated sources.}
  \label{fig:spatial-ebp-exit-lrlw}
\end{figure}

\begin{figure}[t!]
  \centering
  \input{ebp_gexit_46_lrlw_joint_nsw}
  \caption{EBP GEXIT curves of the $(4,6,L,w)$ and
    $(4,6)$ ensembles for transmission over AWGN channels which BSC
    correlation between the sources.}
  \label{fig:spatial-ebp-gexit-lrlw}
\end{figure}


%% file: protograph_jd.tex
\begin{tikzpicture}[scale=0.32]
\useasboundingbox (0,-10) rectangle (18,4);
\begin{scope}[yshift=-15cm] 
  \foreach \x in {2,4,6,8,14,16} {
    \draw[fill=black] (\x,10)+(-5pt,-5pt) rectangle +(5pt,5pt)
    node[left=4pt,below=7pt,outer sep=0pt,inner sep=0pt] (c\x) {};
    \node[outer sep=0pt,inner sep=0pt] (cb\x) at (\x,10) {};
    \foreach \y in {8} {
      \shade[ball color=blue] (\x,\y) circle (6pt) node[outer
      sep=0pt,inner sep=0pt] (vb\x\y) {};
    }	
  }
  \foreach \x in {0,18} {
    \draw[fill=black] (\x,10)+(-5pt,-5pt) rectangle +(5pt,5pt)
    node[left=4pt,below=7pt,outer sep=0pt,inner sep=0pt] (c\x) {};
    \node[outer sep=0pt,inner sep=0pt] (cb\x) at (\x,10) {};	
  }
  \foreach \x in {3,7,11,15} {
    \foreach \y in {12} {
      \draw[fill=white] (\x,\y) circle (6pt) node[outer
      sep=0pt,inner sep=0pt] (vb\x\y) {};
    }	
  }
  \begin{pgfonlayer}{foreground}
    \foreach \x in {10.5,11,11.5} {
      \foreach \y in {8,10,12} {
        \draw[fill=black] (\x,\y) circle (0.5pt);
      }
    }
  \end{pgfonlayer}
  \foreach \x in {10,12} {
    \draw[white] (\x,2)+(-5pt,-5pt) rectangle +(5pt,5pt)
    node[left=4pt,below=7pt,outer sep=0pt,inner sep=0pt] (c\x) {};
    \node[outer sep=0pt,inner sep=0pt] (cb\x) at (\x,10) {};
    \foreach \y in {8,12} {
      \draw[white] (\x,\y) circle (6pt) node[outer
      sep=0pt,inner sep=0pt]
      (vb\x\y) {};
    }
  }
  \draw[white,fill=white] (8.75,8) rectangle +(4.5,4.5);
  
  \begin{pgfonlayer}{background}
    \foreach \x/\y/\z in {2/8/0,2/8/2,2/8/4,2/8/6,4/8/0,4/8/2,4/8/4,4/8/6,3/12/0,3/12/2,3/12/4,3/12/6,
      6/8/4,6/8/6,6/8/8,6/8/10,8/8/4,8/8/6,8/8/8,8/8/10,7/12/4,7/12/6,7/12/8,7/12/10,
      14/8/12,14/8/14,14/8/16,14/8/18,16/8/12,16/8/14,16/8/16,16/8/18,15/12/12,15/12/14,15/12/16,15/12/18,11/12/8,11/12/14,10/8/8,12/8/8,10/8/14,12/8/14} {
      \draw (vb\x\y) -- (cb\z);
    }
  \end{pgfonlayer}
\end{scope}
      \foreach \x in {2,4,6,8,14,16} {
        \draw[fill=black] (\x,2)+(-5pt,-5pt) rectangle +(5pt,5pt)
        node[left=4pt,below=7pt,outer sep=0pt,inner sep=0pt] (c\x) {};
        \node[outer sep=0pt,inner sep=0pt] (c\x) at (\x,2) {};
  		\foreach \y in {4} {
          \shade[ball color=blue] (\x,\y) circle (6pt) node[outer
      	  sep=0pt,inner sep=0pt] (v\x\y) {};
    	}	
      }
      \foreach \x in {0,18} {
        \draw[fill=black] (\x,2)+(-5pt,-5pt) rectangle +(5pt,5pt)
        node[left=4pt,below=7pt,outer sep=0pt,inner sep=0pt] (c\x) {};
        \node[outer sep=0pt,inner sep=0pt] (c\x) at (\x,2) {};	
      }
      \foreach \x in {3,7,11,15} {
    	\foreach \y in {0} {
          \draw[fill=white] (\x,\y) circle (6pt) node[outer
      	  sep=0pt,inner sep=0pt] (v\x\y) {};
    	}	
      }
      \begin{pgfonlayer}{foreground}
        \foreach \x in {10.5,11,11.5} {
          \foreach \y in {-1.5,0,2,4} {
            \draw[fill=black] (\x,\y) circle (0.5pt);
          }
        }
      \end{pgfonlayer}
      \foreach \x in {10,12} {
        \draw[white] (\x,2)+(-5pt,-5pt) rectangle +(5pt,5pt)
        node[left=4pt,below=7pt,outer sep=0pt,inner sep=0pt] (c\x) {};
        \node[outer sep=0pt,inner sep=0pt] (c\x) at (\x,2) {};
        \foreach \y in {0,4} {
          \draw[white] (\x,\y) circle (6pt) node[outer
          sep=0pt,inner sep=0pt]
          (v\x\y) {};
        }
      }
      \draw[white,fill=white] (8.75,-0.5) rectangle +(4.5,4.5);
      
      \begin{pgfonlayer}{background}
        \foreach \x/\y/\z in {2/4/0,2/4/2,2/4/4,2/4/6,4/4/0,4/4/2,4/4/4,4/4/6,3/0/0,3/0/2,3/0/4,3/0/6,
          6/4/4,6/4/6,6/4/8,6/4/10,8/4/4,8/4/6,8/4/8,8/4/10,7/0/4,7/0/6,7/0/8,7/0/10,
          14/4/12,14/4/14,14/4/16,14/4/18,16/4/12,16/4/14,16/4/16,16/4/18,15/0/12,15/0/14,15/0/16,15/0/18,11/0/8,11/0/14,10/4/8,12/4/8,10/4/14,12/4/14} {
          \draw (v\x\y) -- (c\z);
        }


\foreach \x in {3,7,15} {
\draw[thick] (\x,0) -- (\x,-3);
      \draw[fill=red] (\x,-1.5)+(-5pt,-5pt) rectangle +(5pt,5pt)
        node[left=4pt,below=7pt,outer sep=0pt,inner sep=0pt] (c\x) {};
}
      \end{pgfonlayer}
  \draw
  [gray,thick,decorate,decoration={brace,amplitude=5pt}]
  (18,-9) -- (0,-9)
  node [black,midway,below=3pt] {$2L+1$};

\end{tikzpicture}


%% file: ebp_exit_46_lrlw_joint_nsw.tex
%

\begin{tikzpicture}[scale=0.4,>=stealth]

\definecolor{mycolor1}{rgb}{0,1,1}
\definecolor{mycolor2}{rgb}{1,0,1}

\begin{axis}[
scale only axis,
width=5in,
height=5in,
xmin=0, xmax=1,
ymin=0, ymax=1,
xlabel=\mbox{\Large $\epsilon$},
ylabel=\mbox{\Large $h^{\text{EBP}}$},
legend style={at={(0.41,.99)}},
xmajorgrids,
ymajorgrids]
\addplot [
color=blue,
line width=1.25pt,
solid
]
coordinates{
 (1,1)
 (0.9985,1)
 (0.997,1)
 (0.9955,1)
 (0.994,1)
 (0.9925,1)
 (0.991,1)
 (0.9895,1)
 (0.988,1)
 (0.9865,1)
 (0.985,1)
 (0.9835,1)
 (0.982,1)
 (0.9805,1)
 (0.979,1)
 (0.9775,1)
 (0.976,1)
 (0.9745,1)
 (0.973,1)
 (0.9715,1)
 (0.97,1)
 (0.9685,1)
 (0.967,1)
 (0.9655,1)
 (0.964,1)
 (0.9625,1)
 (0.961,1)
 (0.9595,1)
 (0.958,1)
 (0.9565,1)
 (0.955,1)
 (0.9535,1)
 (0.952,1)
 (0.9505,1)
 (0.949,1)
 (0.9475,1)
 (0.946,1)
 (0.9445,1)
 (0.943,1)
 (0.9415,1)
 (0.940001,1)
 (0.938501,1)
 (0.937001,0.999999)
 (0.935501,0.999999)
 (0.934001,0.999999)
 (0.932501,0.999999)
 (0.931001,0.999999)
 (0.929501,0.999999)
 (0.928001,0.999999)
 (0.926501,0.999999)
 (0.925002,0.999999)
 (0.923502,0.999999)
 (0.922002,0.999998)
 (0.920502,0.999998)
 (0.919002,0.999998)
 (0.917503,0.999998)
 (0.916003,0.999998)
 (0.914503,0.999998)
 (0.913003,0.999997)
 (0.911504,0.999997)
 (0.910004,0.999997)
 (0.908504,0.999997)
 (0.907005,0.999996)
 (0.905505,0.999996)
 (0.904006,0.999996)
 (0.902506,0.999995)
 (0.901007,0.999995)
 (0.899507,0.999995)
 (0.898008,0.999994)
 (0.896508,0.999994)
 (0.895009,0.999993)
 (0.893509,0.999993)
 (0.89201,0.999992)
 (0.890511,0.999992)
 (0.889011,0.999991)
 (0.887512,0.999991)
 (0.886013,0.99999)
 (0.884514,0.999989)
 (0.883015,0.999988)
 (0.881516,0.999988)
 (0.880017,0.999987)
 (0.878518,0.999986)
 (0.877019,0.999985)
 (0.87552,0.999984)
 (0.874021,0.999983)
 (0.872523,0.999982)
 (0.871024,0.999981)
 (0.869525,0.99998)
 (0.868027,0.999979)
 (0.866528,0.999978)
 (0.86503,0.999976)
 (0.863532,0.999975)
 (0.862034,0.999974)
 (0.860535,0.999972)
 (0.859037,0.999971)
 (0.857539,0.999969)
 (0.856041,0.999967)
 (0.854543,0.999966)
 (0.853046,0.999964)
 (0.851548,0.999962)
 (0.850051,0.99996)
 (0.848553,0.999958)
 (0.847056,0.999956)
 (0.845558,0.999954)
 (0.844061,0.999951)
 (0.842564,0.999949)
 (0.841067,0.999946)
 (0.83957,0.999944)
 (0.838074,0.999941)
 (0.836577,0.999938)
 (0.835081,0.999936)
 (0.833584,0.999933)
 (0.832088,0.99993)
 (0.830592,0.999926)
 (0.829096,0.999923)
 (0.8276,0.99992)
 (0.826105,0.999916)
 (0.824609,0.999912)
 (0.823114,0.999908)
 (0.821618,0.999905)
 (0.820123,0.9999)
 (0.818629,0.999896)
 (0.817134,0.999892)
 (0.815639,0.999887)
 (0.814145,0.999883)
 (0.812651,0.999878)
 (0.811157,0.999873)
 (0.809663,0.999868)
 (0.808169,0.999863)
 (0.806676,0.999857)
 (0.805182,0.999851)
 (0.803689,0.999846)
 (0.802197,0.99984)
 (0.800704,0.999834)
 (0.799212,0.999827)
 (0.797719,0.999821)
 (0.796227,0.999814)
 (0.794736,0.999807)
 (0.793244,0.9998)
 (0.791753,0.999792)
 (0.790262,0.999785)
 (0.788771,0.999777)
 (0.787281,0.999769)
 (0.78579,0.999761)
 (0.7843,0.999752)
 (0.782811,0.999744)
 (0.781321,0.999735)
 (0.779832,0.999725)
 (0.778343,0.999716)
 (0.776855,0.999706)
 (0.775366,0.999696)
 (0.773878,0.999686)
 (0.772391,0.999675)
 (0.770903,0.999665)
 (0.769416,0.999654)
 (0.76793,0.999642)
 (0.766443,0.99963)
 (0.764957,0.999618)
 (0.763472,0.999606)
 (0.761986,0.999594)
 (0.760501,0.999581)
 (0.759017,0.999567)
 (0.757532,0.999554)
 (0.756049,0.99954)
 (0.754565,0.999526)
 (0.753082,0.999511)
 (0.751599,0.999496)
 (0.750117,0.999481)
 (0.748635,0.999465)
 (0.747153,0.999449)
 (0.745672,0.999432)
 (0.744192,0.999415)
 (0.742712,0.999398)
 (0.741232,0.99938)
 (0.739752,0.999362)
 (0.738274,0.999344)
 (0.736795,0.999325)
 (0.735317,0.999305)
 (0.73384,0.999285)
 (0.732363,0.999265)
 (0.730886,0.999244)
 (0.72941,0.999223)
 (0.727935,0.999201)
 (0.72646,0.999179)
 (0.724986,0.999157)
 (0.723512,0.999133)
 (0.722038,0.99911)
 (0.720566,0.999086)
 (0.719093,0.999061)
 (0.717622,0.999036)
 (0.716151,0.99901)
 (0.71468,0.998984)
 (0.71321,0.998957)
 (0.711741,0.998929)
 (0.710272,0.998901)
 (0.708804,0.998873)
 (0.707336,0.998843)
 (0.705869,0.998814)
 (0.704403,0.998783)
 (0.702938,0.998752)
 (0.701473,0.998721)
 (0.700008,0.998688)
 (0.698545,0.998655)
 (0.697082,0.998622)
 (0.69562,0.998588)
 (0.694158,0.998553)
 (0.692697,0.998517)
 (0.691237,0.998481)
 (0.689778,0.998444)
 (0.688319,0.998406)
 (0.686861,0.998367)
 (0.685404,0.998328)
 (0.683948,0.998288)
 (0.682492,0.998247)
 (0.681037,0.998206)
 (0.679583,0.998164)
 (0.67813,0.998121)
 (0.676678,0.998077)
 (0.675226,0.998032)
 (0.673776,0.997986)
 (0.672326,0.99794)
 (0.670877,0.997893)
 (0.669429,0.997845)
 (0.667981,0.997796)
 (0.666535,0.997746)
 (0.66509,0.997695)
 (0.663645,0.997644)
 (0.662201,0.997591)
 (0.660759,0.997538)
 (0.659317,0.997483)
 (0.657876,0.997428)
 (0.656436,0.997372)
 (0.654997,0.997314)
 (0.653559,0.997256)
 (0.652123,0.997197)
 (0.650687,0.997136)
 (0.649252,0.997075)
 (0.647818,0.997012)
 (0.646385,0.996949)
 (0.644954,0.996884)
 (0.643523,0.996819)
 (0.642093,0.996752)
 (0.640665,0.996684)
 (0.639237,0.996615)
 (0.637811,0.996545)
 (0.636386,0.996474)
 (0.634962,0.996401)
 (0.633539,0.996328)
 (0.632117,0.996253)
 (0.630697,0.996177)
 (0.629277,0.996099)
 (0.627859,0.996021)
 (0.626442,0.995941)
 (0.625027,0.99586)
 (0.623612,0.995778)
 (0.622199,0.995694)
 (0.620787,0.995609)
 (0.619376,0.995523)
 (0.617967,0.995435)
 (0.616559,0.995346)
 (0.615152,0.995256)
 (0.613746,0.995164)
 (0.612342,0.995071)
 (0.61094,0.994976)
 (0.609538,0.99488)
 (0.608138,0.994783)
 (0.606739,0.994684)
 (0.605342,0.994583)
 (0.603946,0.994481)
 (0.602552,0.994378)
 (0.601159,0.994273)
 (0.599767,0.994166)
 (0.598377,0.994058)
 (0.596989,0.993948)
 (0.595602,0.993837)
 (0.594216,0.993724)
 (0.592832,0.993609)
 (0.59145,0.993493)
 (0.590069,0.993375)
 (0.58869,0.993255)
 (0.587312,0.993134)
 (0.585936,0.99301)
 (0.584561,0.992886)
 (0.583188,0.992759)
 (0.581817,0.99263)
 (0.580448,0.9925)
 (0.57908,0.992368)
 (0.577714,0.992234)
 (0.576349,0.992098)
 (0.574986,0.99196)
 (0.573625,0.991821)
 (0.572266,0.991679)
 (0.570908,0.991536)
 (0.569552,0.99139)
 (0.568198,0.991243)
 (0.566846,0.991093)
 (0.565496,0.990942)
 (0.564147,0.990788)
 (0.562801,0.990633)
 (0.561456,0.990475)
 (0.560113,0.990315)
 (0.558772,0.990153)
 (0.557433,0.989989)
 (0.556096,0.989823)
 (0.554761,0.989655)
 (0.553427,0.989484)
 (0.552096,0.989311)
 (0.550767,0.989136)
 (0.54944,0.988959)
 (0.548114,0.988779)
 (0.546791,0.988597)
 (0.54547,0.988413)
 (0.544151,0.988226)
 (0.542834,0.988037)
 (0.541519,0.987846)
 (0.540207,0.987652)
 (0.538896,0.987456)
 (0.537588,0.987257)
 (0.536281,0.987056)
 (0.534977,0.986852)
 (0.533676,0.986646)
 (0.532376,0.986437)
 (0.531079,0.986225)
 (0.529784,0.986011)
 (0.528491,0.985794)
 (0.5272,0.985575)
 (0.525912,0.985353)
 (0.524626,0.985128)
 (0.523343,0.984901)
 (0.522062,0.984671)
 (0.520783,0.984437)
 (0.519507,0.984202)
 (0.518233,0.983963)
 (0.516961,0.983722)
 (0.515692,0.983477)
 (0.514426,0.98323)
 (0.513162,0.98298)
 (0.5119,0.982727)
 (0.510641,0.982471)
 (0.509384,0.982211)
 (0.508131,0.981949)
 (0.506879,0.981684)
 (0.50563,0.981416)
 (0.504384,0.981145)
 (0.503141,0.98087)
 (0.5019,0.980592)
 (0.500662,0.980312)
 (0.499426,0.980028)
 (0.498194,0.97974)
 (0.496964,0.97945)
 (0.495736,0.979156)
 (0.494512,0.978859)
 (0.49329,0.978559)
 (0.492071,0.978255)
 (0.490855,0.977948)
 (0.489642,0.977637)
 (0.488432,0.977323)
 (0.487224,0.977006)
 (0.486019,0.976685)
 (0.484818,0.97636)
 (0.483619,0.976032)
 (0.482423,0.9757)
 (0.48123,0.975365)
 (0.480041,0.975026)
 (0.478854,0.974684)
 (0.47767,0.974337)
 (0.476489,0.973987)
 (0.475312,0.973634)
 (0.474137,0.973276)
 (0.472966,0.972915)
 (0.471797,0.97255)
 (0.470632,0.972181)
 (0.46947,0.971808)
 (0.468312,0.971431)
 (0.467156,0.97105)
 (0.466004,0.970665)
 (0.464855,0.970276)
 (0.463709,0.969884)
 (0.462566,0.969487)
 (0.461427,0.969086)
 (0.460291,0.968681)
 (0.459159,0.968271)
 (0.45803,0.967858)
 (0.456904,0.96744)
 (0.455782,0.967018)
 (0.454663,0.966592)
 (0.453548,0.966162)
 (0.452436,0.965727)
 (0.451327,0.965287)
 (0.450223,0.964844)
 (0.449121,0.964396)
 (0.448024,0.963943)
 (0.446929,0.963486)
 (0.445839,0.963025)
 (0.444752,0.962559)
 (0.443669,0.962088)
 (0.442589,0.961613)
 (0.441513,0.961133)
 (0.440441,0.960648)
 (0.439373,0.960159)
 (0.438308,0.959665)
 (0.437247,0.959166)
 (0.43619,0.958662)
 (0.435137,0.958154)
 (0.434087,0.957641)
 (0.433042,0.957122)
 (0.432,0.956599)
 (0.430962,0.956071)
 (0.429928,0.955538)
 (0.428898,0.955)
 (0.427872,0.954457)
 (0.42685,0.953908)
 (0.425833,0.953355)
 (0.424819,0.952796)
 (0.423809,0.952233)
 (0.422803,0.951664)
 (0.421802,0.951089)
 (0.420804,0.95051)
 (0.419811,0.949925)
 (0.418822,0.949335)
 (0.417837,0.948739)
 (0.416856,0.948138)
 (0.41588,0.947532)
 (0.414908,0.94692)
 (0.41394,0.946303)
 (0.412976,0.94568)
 (0.412017,0.945051)
 (0.411062,0.944417)
 (0.410112,0.943777)
 (0.409166,0.943132)
 (0.408224,0.942481)
 (0.407287,0.941824)
 (0.406354,0.941161)
 (0.405426,0.940493)
 (0.404502,0.939818)
 (0.403583,0.939138)
 (0.402668,0.938452)
 (0.401758,0.93776)
 (0.400853,0.937062)
 (0.399952,0.936358)
 (0.399056,0.935647)
 (0.398165,0.934931)
 (0.397278,0.934209)
 (0.396396,0.933481)
 (0.395519,0.932746)
 (0.394647,0.932005)
 (0.393779,0.931258)
 (0.392917,0.930505)
 (0.392059,0.929745)
 (0.391206,0.928979)
 (0.390358,0.928207)
 (0.389515,0.927428)
 (0.388677,0.926643)
 (0.387843,0.925851)
 (0.387015,0.925053)
 (0.386192,0.924248)
 (0.385374,0.923437)
 (0.384561,0.922619)
 (0.383753,0.921794)
 (0.38295,0.920963)
 (0.382152,0.920125)
 (0.38136,0.919281)
 (0.380573,0.918429)
 (0.379791,0.917571)
 (0.379014,0.916706)
 (0.378242,0.915834)
 (0.377476,0.914955)
 (0.376715,0.914069)
 (0.37596,0.913177)
 (0.375209,0.912277)
 (0.374465,0.91137)
 (0.373725,0.910456)
 (0.372991,0.909535)
 (0.372263,0.908607)
 (0.37154,0.907672)
 (0.370823,0.906729)
 (0.370111,0.90578)
 (0.369405,0.904823)
 (0.368704,0.903859)
 (0.368009,0.902887)
 (0.36732,0.901908)
 (0.366636,0.900922)
 (0.365958,0.899928)
 (0.365286,0.898927)
 (0.36462,0.897918)
 (0.363959,0.896902)
 (0.363304,0.895879)
 (0.362655,0.894847)
 (0.362012,0.893809)
 (0.361375,0.892762)
 (0.360744,0.891708)
 (0.360118,0.890646)
 (0.359499,0.889577)
 (0.358886,0.888499)
 (0.358278,0.887414)
 (0.357677,0.886321)
 (0.357082,0.88522)
 (0.356493,0.884112)
 (0.35591,0.882995)
 (0.355333,0.881871)
 (0.354763,0.880738)
 (0.354198,0.879598)
 (0.35364,0.878449)
 (0.353088,0.877293)
 (0.352543,0.876128)
 (0.352004,0.874956)
 (0.351471,0.873775)
 (0.350944,0.872586)
 (0.350424,0.871389)
 (0.349911,0.870183)
 (0.349404,0.86897)
 (0.348903,0.867748)
 (0.348409,0.866517)
 (0.347922,0.865279)
 (0.347441,0.864032)
 (0.346966,0.862777)
 (0.346499,0.861513)
 (0.346038,0.860241)
 (0.345583,0.85896)
 (0.345136,0.857671)
 (0.344695,0.856374)
 (0.344261,0.855068)
 (0.343834,0.853753)
 (0.343413,0.85243)
 (0.343,0.851098)
 (0.342593,0.849758)
 (0.342193,0.848409)
 (0.341801,0.847051)
 (0.341415,0.845685)
 (0.341036,0.844309)
 (0.340665,0.842926)
 (0.3403,0.841533)
 (0.339943,0.840131)
 (0.339592,0.838721)
 (0.339249,0.837302)
 (0.338913,0.835874)
 (0.338584,0.834437)
 (0.338263,0.832992)
 (0.337949,0.831537)
 (0.337642,0.830074)
 (0.337342,0.828601)
 (0.33705,0.82712)
 (0.336766,0.825629)
 (0.336488,0.82413)
 (0.336219,0.822621)
 (0.335956,0.821104)
 (0.335702,0.819577)
 (0.335454,0.818042)
 (0.335215,0.816497)
 (0.334983,0.814943)
 (0.334759,0.81338)
 (0.334542,0.811808)
 (0.334333,0.810227)
 (0.334132,0.808637)
 (0.333939,0.807037)
 (0.333754,0.805429)
 (0.333576,0.803811)
 (0.333407,0.802184)
 (0.333245,0.800547)
 (0.333091,0.798902)
 (0.332945,0.797247)
 (0.332807,0.795583)
 (0.332678,0.793909)
 (0.332556,0.792226)
 (0.332443,0.790534)
 (0.332338,0.788833)
 (0.33224,0.787123)
 (0.332152,0.785403)
 (0.332071,0.783673)
 (0.331999,0.781935)
 (0.331935,0.780187)
 (0.331879,0.77843)
 (0.331832,0.776663)
 (0.331793,0.774887)
 (0.331763,0.773102)
 (0.331741,0.771307)
 (0.331728,0.769503)
 (0.331724,0.76769)
 (0.331728,0.765867)
 (0.331741,0.764035)
 (0.331762,0.762193)
 (0.331792,0.760342)
 (0.331831,0.758482)
 (0.331879,0.756612)
 (0.331935,0.754733)
 (0.332001,0.752845)
 (0.332075,0.750947)
 (0.332159,0.74904)
 (0.332251,0.747123)
 (0.332353,0.745197)
 (0.332463,0.743262)
 (0.332583,0.741317)
 (0.332712,0.739363)
 (0.33285,0.7374)
 (0.332997,0.735427)
 (0.333154,0.733445)
 (0.333319,0.731454)
 (0.333495,0.729453)
 (0.333679,0.727443)
 (0.333873,0.725424)
 (0.334077,0.723396)
 (0.33429,0.721358)
 (0.334513,0.719311)
 (0.334745,0.717254)
 (0.334987,0.715189)
 (0.335239,0.713114)
 (0.3355,0.71103)
 (0.335772,0.708937)
 (0.336053,0.706835)
 (0.336344,0.704724)
 (0.336645,0.702603)
 (0.336956,0.700473)
 (0.337277,0.698335)
 (0.337608,0.696187)
 (0.337949,0.69403)
 (0.3383,0.691864)
 (0.338662,0.689689)
 (0.339033,0.687505)
 (0.339416,0.685312)
 (0.339808,0.68311)
 (0.340211,0.6809)
 (0.340624,0.67868)
 (0.341048,0.676452)
 (0.341482,0.674214)
 (0.341928,0.671968)
 (0.342383,0.669713)
 (0.34285,0.66745)
 (0.343327,0.665177)
 (0.343815,0.662896)
 (0.344314,0.660607)
 (0.344824,0.658308)
 (0.345344,0.656001)
 (0.345876,0.653686)
 (0.346419,0.651362)
 (0.346974,0.64903)
 (0.347539,0.646689)
 (0.348116,0.64434)
 (0.348704,0.641982)
 (0.349303,0.639616)
 (0.349914,0.637242)
 (0.350536,0.63486)
 (0.35117,0.632469)
 (0.351816,0.63007)
 (0.352473,0.627664)
 (0.353142,0.625249)
 (0.353823,0.622826)
 (0.354516,0.620395)
 (0.355221,0.617956)
 (0.355937,0.61551)
 (0.356666,0.613055)
 (0.357407,0.610593)
 (0.358161,0.608123)
 (0.358926,0.605646)
 (0.359704,0.603161)
 (0.360495,0.600668)
 (0.361298,0.598168)
 (0.362113,0.59566)
 (0.362941,0.593145)
 (0.363782,0.590623)
 (0.364636,0.588094)
 (0.365503,0.585557)
 (0.366382,0.583013)
 (0.367275,0.580462)
 (0.368181,0.577904)
 (0.3691,0.575339)
 (0.370032,0.572768)
 (0.370978,0.570189)
 (0.371938,0.567604)
 (0.37291,0.565012)
 (0.373897,0.562413)
 (0.374897,0.559808)
 (0.375911,0.557197)
 (0.376939,0.554579)
 (0.377981,0.551955)
 (0.379037,0.549324)
 (0.380107,0.546688)
 (0.381192,0.544045)
 (0.382291,0.541396)
 (0.383404,0.538742)
 (0.384532,0.536081)
 (0.385675,0.533415)
 (0.386832,0.530743)
 (0.388005,0.528065)
 (0.389192,0.525382)
 (0.390395,0.522694)
 (0.391613,0.52)
 (0.392846,0.517301)
 (0.394094,0.514597)
 (0.395359,0.511887)
 (0.396638,0.509173)
 (0.397934,0.506454)
 (0.399246,0.50373)
 (0.400573,0.501001)
 (0.401917,0.498268)
 (0.403277,0.49553)
 (0.404654,0.492787)
 (0.406047,0.490041)
 (0.407457,0.48729)
 (0.408884,0.484535)
 (0.410328,0.481776)
 (0.411788,0.479013)
 (0.413267,0.476246)
 (0.414762,0.473475)
 (0.416275,0.470701)
 (0.417806,0.467924)
 (0.419355,0.465143)
 (0.420921,0.462358)
 (0.422506,0.459571)
 (0.424109,0.45678)
 (0.425731,0.453986)
 (0.427371,0.45119)
 (0.429031,0.44839)
 (0.430709,0.445589)
 (0.432406,0.442784)
 (0.434123,0.439977)
 (0.435859,0.437168)
 (0.437615,0.434357)
 (0.439391,0.431544)
 (0.441187,0.428728)
 (0.443003,0.425911)
 (0.44484,0.423092)
 (0.446698,0.420272)
 (0.448576,0.41745)
 (0.450476,0.414627)
 (0.452397,0.411803)
 (0.454339,0.408977)
 (0.456304,0.406151)
 (0.45829,0.403324)
 (0.460299,0.400496)
 (0.46233,0.397668)
 (0.464383,0.394839)
 (0.46646,0.39201)
 (0.46856,0.38918)
 (0.470683,0.386351)
 (0.47283,0.383522)
 (0.475001,0.380693)
 (0.477197,0.377864)
 (0.479416,0.375036)
 (0.481661,0.372208)
 (0.483931,0.369382)
 (0.486226,0.366556)
 (0.488547,0.363731)
 (0.490893,0.360908)
 (0.493267,0.358085)
 (0.495666,0.355265)
 (0.498093,0.352446)
 (0.500547,0.349628)
 (0.503028,0.346813)
 (0.505537,0.344)
 (0.508075,0.341188)
 (0.510641,0.33838)
 (0.513236,0.335573)
 (0.515861,0.33277)
 (0.518515,0.329969)
 (0.5212,0.327171)
 (0.523915,0.324376)
 (0.526661,0.321585)
 (0.529438,0.318796)
 (0.532247,0.316012)
 (0.535088,0.313231)
 (0.537962,0.310454)
 (0.540868,0.307681)
 (0.543809,0.304912)
 (0.546783,0.302147)
 (0.549792,0.299387)
 (0.552836,0.296631)
 (0.555915,0.293881)
 (0.55903,0.291135)
 (0.562181,0.288394)
 (0.56537,0.285658)
 (0.568596,0.282928)
 (0.57186,0.280204)
 (0.575163,0.277485)
 (0.578505,0.274772)
 (0.581887,0.272065)
 (0.585309,0.269364)
 (0.588772,0.266669)
 (0.592277,0.263981)
 (0.595824,0.2613)
 (0.599415,0.258626)
 (0.603049,0.255958)
 (0.606727,0.253298)
 (0.61045,0.250644)
 (0.61422,0.247999)
 (0.618035,0.245361)
 (0.621898,0.24273)
 (0.625809,0.240108)
 (0.629769,0.237494)
 (0.633779,0.234888)
 (0.637839,0.23229)
 (0.641951,0.229701)
 (0.646114,0.227121)
 (0.650331,0.224549)
 (0.654602,0.221987)
 (0.658928,0.219434)
 (0.66331,0.21689)
 (0.667749,0.214356)
 (0.672245,0.211831)
 (0.676801,0.209316)
 (0.681417,0.206811)
 (0.686094,0.204317)
 (0.690833,0.201832)
 (0.695635,0.199359)
 (0.700502,0.196895)
 (0.705435,0.194443)
 (0.710435,0.192001)
 (0.715503,0.189571)
 (0.72064,0.187151)
 (0.725849,0.184743)
 (0.73113,0.182347)
 (0.736484,0.179962)
 (0.741913,0.17759)
 (0.747419,0.175229)
 (0.753003,0.17288)
 (0.758667,0.170544)
 (0.764412,0.16822)
 (0.770239,0.165908)
 (0.776151,0.163609)
 (0.78215,0.161323)
 (0.788236,0.159051)
 (0.794412,0.156791)
 (0.80068,0.154545)
 (0.807042,0.152312)
 (0.813499,0.150092)
 (0.820054,0.147886)
 (0.826709,0.145695)
 (0.833465,0.143517)
 (0.840325,0.141353)
 (0.847292,0.139203)
 (0.854368,0.137068)
 (0.861555,0.134948)
 (0.868855,0.132842)
 (0.876271,0.13075)
 (0.883806,0.128674)
 (0.891463,0.126613)
 (0.899244,0.124567)
 (0.907152,0.122536)
 (0.91519,0.12052)
 (0.923361,0.11852)
 (0.931668,0.116536)
 (0.940114,0.114567)
 (0.948704,0.112614)
 (0.957439,0.110677)
 (0.966325,0.108756)
 (0.975363,0.106851)
 (0.984559,0.104963)
 (0.993916,0.103091)
 (1.00344,0.101235)

};
\addlegendentry{\Large JIT - punc $(4,6)$}

\addplot [
color=black,
fill=gray,fill opacity=0.4,
solid
]
coordinates{
 (1,1)
 (0.9985,1)
 (0.997,1)
 (0.9955,1)
 (0.994,1)
 (0.9925,1)
 (0.991,1)
 (0.9895,1)
 (0.988,1)
 (0.9865,1)
 (0.985,1)
 (0.9835,1)
 (0.982,1)
 (0.9805,1)
 (0.979,1)
 (0.9775,1)
 (0.976,1)
 (0.9745,1)
 (0.973,1)
 (0.9715,1)
 (0.97,1)
 (0.9685,1)
 (0.967,1)
 (0.9655,1)
 (0.964,1)
 (0.9625,1)
 (0.961,1)
 (0.9595,1)
 (0.958,1)
 (0.9565,1)
 (0.955,1)
 (0.9535,1)
 (0.952,1)
 (0.9505,1)
 (0.949,1)
 (0.9475,1)
 (0.946,1)
 (0.9445,1)
 (0.943,1)
 (0.9415,1)
 (0.940001,1)
 (0.938501,1)
 (0.937001,0.999999)
 (0.935501,0.999999)
 (0.934001,0.999999)
 (0.932501,0.999999)
 (0.931001,0.999999)
 (0.929501,0.999999)
 (0.928001,0.999999)
 (0.926501,0.999999)
 (0.925002,0.999999)
 (0.923502,0.999999)
 (0.922002,0.999998)
 (0.920502,0.999998)
 (0.919002,0.999998)
 (0.917503,0.999998)
 (0.916003,0.999998)
 (0.914503,0.999998)
 (0.913003,0.999997)
 (0.911504,0.999997)
 (0.910004,0.999997)
 (0.908504,0.999997)
 (0.907005,0.999996)
 (0.905505,0.999996)
 (0.904006,0.999996)
 (0.902506,0.999995)
 (0.901007,0.999995)
 (0.899507,0.999995)
 (0.898008,0.999994)
 (0.896508,0.999994)
 (0.895009,0.999993)
 (0.893509,0.999993)
 (0.89201,0.999992)
 (0.890511,0.999992)
 (0.889011,0.999991)
 (0.887512,0.999991)
 (0.886013,0.99999)
 (0.884514,0.999989)
 (0.883015,0.999988)
 (0.881516,0.999988)
 (0.880017,0.999987)
 (0.878518,0.999986)
 (0.877019,0.999985)
 (0.87552,0.999984)
 (0.874021,0.999983)
 (0.872523,0.999982)
 (0.871024,0.999981)
 (0.869525,0.99998)
 (0.868027,0.999979)
 (0.866528,0.999978)
 (0.86503,0.999976)
 (0.863532,0.999975)
 (0.862034,0.999974)
 (0.860535,0.999972)
 (0.859037,0.999971)
 (0.857539,0.999969)
 (0.856041,0.999967)
 (0.854543,0.999966)
 (0.853046,0.999964)
 (0.851548,0.999962)
 (0.850051,0.99996)
 (0.848553,0.999958)
 (0.847056,0.999956)
 (0.845558,0.999954)
 (0.844061,0.999951)
 (0.842564,0.999949)
 (0.841067,0.999946)
 (0.83957,0.999944)
 (0.838074,0.999941)
 (0.836577,0.999938)
 (0.835081,0.999936)
 (0.833584,0.999933)
 (0.832088,0.99993)
 (0.830592,0.999926)
 (0.829096,0.999923)
 (0.8276,0.99992)
 (0.826105,0.999916)
 (0.824609,0.999912)
 (0.823114,0.999908)
 (0.821618,0.999905)
 (0.820123,0.9999)
 (0.818629,0.999896)
 (0.817134,0.999892)
 (0.815639,0.999887)
 (0.814145,0.999883)
 (0.812651,0.999878)
 (0.811157,0.999873)
 (0.809663,0.999868)
 (0.808169,0.999863)
 (0.806676,0.999857)
 (0.805182,0.999851)
 (0.803689,0.999846)
 (0.802197,0.99984)
 (0.800704,0.999834)
 (0.799212,0.999827)
 (0.797719,0.999821)
 (0.796227,0.999814)
 (0.794736,0.999807)
 (0.793244,0.9998)
 (0.791753,0.999792)
 (0.790262,0.999785)
 (0.788771,0.999777)
 (0.787281,0.999769)
 (0.78579,0.999761)
 (0.7843,0.999752)
 (0.782811,0.999744)
 (0.781321,0.999735)
 (0.779832,0.999725)
 (0.778343,0.999716)
 (0.776855,0.999706)
 (0.775366,0.999696)
 (0.773878,0.999686)
 (0.772391,0.999675)
 (0.770903,0.999665)
 (0.769416,0.999654)
 (0.76793,0.999642)
 (0.766443,0.99963)
 (0.764957,0.999618)
 (0.763472,0.999606)
 (0.761986,0.999594)
 (0.760501,0.999581)
 (0.759017,0.999567)
 (0.757532,0.999554)
 (0.756049,0.99954)
 (0.754565,0.999526)
 (0.753082,0.999511)
 (0.751599,0.999496)
 (0.750117,0.999481)
 (0.748635,0.999465)
 (0.747153,0.999449)
 (0.745672,0.999432)
 (0.744192,0.999415)
 (0.742712,0.999398)
 (0.741232,0.99938)
 (0.739752,0.999362)
 (0.738274,0.999344)
 (0.736795,0.999325)
 (0.735317,0.999305)
 (0.73384,0.999285)
 (0.732363,0.999265)
 (0.730886,0.999244)
 (0.72941,0.999223)
 (0.727935,0.999201)
 (0.72646,0.999179)
 (0.724986,0.999157)
 (0.723512,0.999133)
 (0.722038,0.99911)
 (0.720566,0.999086)
 (0.719093,0.999061)
 (0.717622,0.999036)
 (0.716151,0.99901)
 (0.71468,0.998984)
 (0.71321,0.998957)
 (0.711741,0.998929)
 (0.710272,0.998901)
 (0.708804,0.998873)
 (0.707336,0.998843)
 (0.705869,0.998814)
 (0.704403,0.998783)
 (0.702938,0.998752)
 (0.701473,0.998721)
 (0.700008,0.998688)
 (0.698545,0.998655)
 (0.697082,0.998622)
 (0.69562,0.998588)
 (0.694158,0.998553)
 (0.692697,0.998517)
 (0.691237,0.998481)
 (0.689778,0.998444)
 (0.688319,0.998406)
 (0.686861,0.998367)
 (0.685404,0.998328)
 (0.683948,0.998288)
 (0.682492,0.998247)
 (0.681037,0.998206)
 (0.679583,0.998164)
 (0.67813,0.998121)
 (0.676678,0.998077)
 (0.675226,0.998032)
 (0.673776,0.997986)
 (0.672326,0.99794)
 (0.670877,0.997893)
 (0.669429,0.997845)
 (0.667981,0.997796)
 (0.666535,0.997746)
 (0.66509,0.997695)
 (0.663645,0.997644)
 (0.662201,0.997591)
 (0.660759,0.997538)
 (0.659317,0.997483)
 (0.657876,0.997428)
 (0.656436,0.997372)
 (0.654997,0.997314)
 (0.653559,0.997256)
 (0.652123,0.997197)
 (0.650687,0.997136)
 (0.649252,0.997075)
 (0.647818,0.997012)
 (0.646385,0.996949)
 (0.644954,0.996884)
 (0.643523,0.996819)
 (0.642093,0.996752)
 (0.640665,0.996684)
 (0.639237,0.996615)
 (0.637811,0.996545)
 (0.636386,0.996474)
 (0.634962,0.996401)
 (0.633539,0.996328)
 (0.632117,0.996253)
 (0.630697,0.996177)
 (0.629277,0.996099)
 (0.627859,0.996021)
 (0.626442,0.995941)
 (0.625027,0.99586)
 (0.625027,0)

} |- (axis cs:1,0) -- cycle;
\addlegendentry{\Large MAP - punc $(4,6)$}

\addplot [
color=purple,
solid
]
coordinates{
 (1.00723,0.99475)
 (0.99264,0.994405)
 (0.978074,0.994037)
 (0.963534,0.993644)
 (0.949022,0.993224)
 (0.934539,0.992774)
 (0.920089,0.992291)
 (0.905676,0.991771)
 (0.891304,0.99121)
 (0.876978,0.990602)
 (0.862704,0.989942)
 (0.848491,0.989222)
 (0.834346,0.988432)
 (0.820281,0.987563)
 (0.806309,0.986602)
 (0.792446,0.985532)
 (0.778711,0.984334)
 (0.765129,0.982984)
 (0.751731,0.981451)
 (0.73856,0.979693)
 (0.725673,0.977655)
 (0.713155,0.975255)
 (0.701135,0.972372)
 (0.689824,0.968808)
 (0.679588,0.96423)
 (0.670958,0.958176)
 (0.663728,0.950878)
 (0.655864,0.944166)
 (0.646687,0.938585)
 (0.636861,0.933522)
 (0.627038,0.9284)
 (0.617936,0.922601)
 (0.611335,0.914615)
 (0.639119,0.87879)
 (0.631094,0.871882)
 (0.621856,0.865807)
 (0.613653,0.858806)
 (0.613525,0.845397)
 (0.636095,0.81517)
 (0.626791,0.808714)
 (0.617439,0.80217)
 (0.610962,0.793406)
 (0.639976,0.7597)
 (0.63214,0.751833)
 (0.622125,0.745303)
 (0.613346,0.737805)
 (0.614903,0.723258)
 (0.637384,0.695546)
 (0.627527,0.688459)
 (0.61734,0.681434)
 (0.610832,0.671983)
 (0.640281,0.641102)
 (0.633352,0.63191)
 (0.622445,0.624828)
 (0.613014,0.616751)
 (0.616114,0.601515)
 (0.638694,0.576179)
 (0.628409,0.568286)
 (0.617224,0.560685)
 (0.610789,0.550487)
 (0.639228,0.523205)
 (0.634752,0.512155)
 (0.622832,0.50439)
 (0.612657,0.495637)
 (0.616845,0.480249)
 (0.639826,0.457159)
 (0.629479,0.448223)
 (0.617088,0.439921)
 (0.610856,0.428937)
 (0.636166,0.405833)
 (0.636334,0.392629)
 (0.62331,0.384004)
 (0.612282,0.37445)
 (0.617246,0.359253)
 (0.64036,0.338556)
 (0.630798,0.328312)
 (0.616924,0.319134)
 (0.611043,0.307368)
 (0.632425,0.288013)
 (0.638014,0.273408)
 (0.623916,0.263688)
 (0.611902,0.253183)
 (0.617505,0.238375)
 (0.639766,0.220294)
 (0.632436,0.208616)
 (0.616724,0.198319)
 (0.611337,0.18582)
 (0.629578,0.169273)
 (0.639518,0.154561)
 (0.62471,0.143475)
 (0.611536,0.131831)
 (0.617689,0.117563)
 (0.63803,0.102027)
 (0.63452,0.089223)
 (0.616985,0.077422)
 (0.614695,0.0641169)
 (0.642043,0.0490721)
 (0.708426,0.0333345)
 (0.924887,0.0166526)
 (1,0.01)
};
\addlegendentry{\Large JIT - $L = 16,w=2$}

\addplot [
color=mycolor2,
solid
]
coordinates{
 (1.00787,0.983329)
 (0.993476,0.982742)
 (0.979111,0.982126)
 (0.964773,0.981479)
 (0.950466,0.980799)
 (0.936192,0.980082)
 (0.921953,0.979325)
 (0.907754,0.978525)
 (0.893598,0.977678)
 (0.879491,0.976777)
 (0.865438,0.975819)
 (0.851446,0.974794)
 (0.837523,0.973697)
 (0.823678,0.972518)
 (0.809921,0.971247)
 (0.796264,0.969871)
 (0.782721,0.968377)
 (0.769309,0.966749)
 (0.756047,0.964967)
 (0.742957,0.96301)
 (0.730068,0.960851)
 (0.717412,0.958457)
 (0.705031,0.95579)
 (0.692977,0.9528)
 (0.68132,0.949425)
 (0.670151,0.94558)
 (0.659601,0.941153)
 (0.649854,0.935983)
 (0.641192,0.929835)
 (0.634024,0.922374)
 (0.628868,0.913199)
 (0.626021,0.902114)
 (0.624997,0.889565)
 (0.624781,0.876389)
 (0.624744,0.86308)
 (0.624749,0.849742)
 (0.624742,0.836412)
 (0.624747,0.823074)
 (0.624745,0.80974)
 (0.624744,0.796406)
 (0.624748,0.783069)
 (0.624742,0.769738)
 (0.624749,0.756398)
 (0.624742,0.743068)
 (0.624748,0.729729)
 (0.624744,0.716397)
 (0.624746,0.703061)
 (0.624747,0.689726)
 (0.624743,0.676393)
 (0.624749,0.663055)
 (0.624742,0.649724)
 (0.624749,0.636386)
 (0.624744,0.623054)
 (0.624747,0.609717)
 (0.624746,0.596383)
 (0.624744,0.583049)
 (0.624749,0.569712)
 (0.624742,0.55638)
 (0.624749,0.543042)
 (0.624743,0.52971)
 (0.624748,0.516373)
 (0.624745,0.50304)
 (0.624745,0.489705)
 (0.624748,0.476369)
 (0.624743,0.463036)
 (0.624749,0.449699)
 (0.624743,0.436367)
 (0.624749,0.42303)
 (0.624744,0.409697)
 (0.624746,0.396361)
 (0.624747,0.383026)
 (0.624743,0.369692)
 (0.624749,0.356356)
 (0.624742,0.343023)
 (0.624749,0.329686)
 (0.624744,0.316353)
 (0.624747,0.303017)
 (0.624746,0.289683)
 (0.624744,0.276348)
 (0.624749,0.263013)
 (0.624743,0.249679)
 (0.624749,0.236343)
 (0.624743,0.22301)
 (0.624748,0.209674)
 (0.624745,0.196339)
 (0.624745,0.183005)
 (0.624748,0.169669)
 (0.624743,0.156335)
 (0.624751,0.143)
 (0.624748,0.129665)
 (0.624773,0.116327)
 (0.624852,0.102984)
 (0.625228,0.0896145)
 (0.62686,0.0761499)
 (0.633612,0.0623708)
 (0.659394,0.0477237)
 (0.74846,0.0314331)
 (1.04578,0.0143522)

};
\addlegendentry{\Large JIT - $L = 32,w=4$}

\addplot [
color=green,
solid
]
coordinates{
 (1.0006,0.988587)
 (0.985987,0.988203)
 (0.971393,0.987799)
 (0.956819,0.987374)
 (0.942267,0.986925)
 (0.927738,0.98645)
 (0.913235,0.985946)
 (0.898761,0.985411)
 (0.884321,0.984841)
 (0.869918,0.98423)
 (0.855558,0.983575)
 (0.841247,0.982869)
 (0.826991,0.982104)
 (0.812801,0.981273)
 (0.798684,0.980366)
 (0.784654,0.979372)
 (0.770722,0.978277)
 (0.756907,0.977065)
 (0.743226,0.975718)
 (0.729705,0.974211)
 (0.716371,0.972516)
 (0.703265,0.970596)
 (0.690437,0.968403)
 (0.677957,0.965871)
 (0.66593,0.962905)
 (0.654521,0.959357)
 (0.644011,0.95498)
 (0.634931,0.949319)
 (0.628297,0.941525)
 (0.625267,0.930675)
 (0.624731,0.917768)
 (0.62469,0.904465)
 (0.62468,0.891137)
 (0.624674,0.877807)
 (0.62467,0.864474)
 (0.624667,0.851141)
 (0.624666,0.837806)
 (0.624667,0.824469)
 (0.624668,0.811133)
 (0.62467,0.797795)
 (0.624673,0.784458)
 (0.624677,0.771119)
 (0.624681,0.757781)
 (0.624685,0.744443)
 (0.62469,0.731104)
 (0.624694,0.717766)
 (0.624699,0.704428)
 (0.624704,0.691089)
 (0.624708,0.677751)
 (0.624712,0.664414)
 (0.624716,0.651076)
 (0.62472,0.637739)
 (0.624724,0.624402)
 (0.624727,0.611065)
 (0.62473,0.597728)
 (0.624732,0.584392)
 (0.624735,0.571056)
 (0.624737,0.55772)
 (0.624738,0.544384)
 (0.62474,0.531049)
 (0.624741,0.517713)
 (0.624742,0.504378)
 (0.624743,0.491043)
 (0.624744,0.477708)
 (0.624744,0.464372)
 (0.624745,0.451038)
 (0.624745,0.437703)
 (0.624745,0.424368)
 (0.624746,0.411033)
 (0.624746,0.397698)
 (0.624746,0.384363)
 (0.624746,0.371028)
 (0.624746,0.357694)
 (0.624746,0.344359)
 (0.624746,0.331024)
 (0.624746,0.317689)
 (0.624746,0.304354)
 (0.624746,0.29102)
 (0.624746,0.277685)
 (0.624746,0.26435)
 (0.624746,0.251015)
 (0.624746,0.237681)
 (0.624746,0.224346)
 (0.624746,0.211011)
 (0.624746,0.197676)
 (0.624746,0.184341)
 (0.624746,0.171007)
 (0.624746,0.157672)
 (0.624746,0.144337)
 (0.624746,0.131002)
 (0.624746,0.117667)
 (0.624746,0.104333)
 (0.624747,0.0909977)
 (0.62476,0.077662)
 (0.624888,0.0643185)
 (0.626195,0.0509131)
 (0.63825,0.0370912)
 (0.729275,0.0215315)
 (1,0.008)
};
\addlegendentry{\Large JIT - $L = 64,w=5$}

\addplot [
color=blue,
line width=1.25pt,
dash pattern=on 1pt off 3pt on 3pt off 3pt
]
coordinates{
 (0.331724,0)
 (0.331724,0.76769)

};

\end{axis}

\end{tikzpicture}


%% file: ebp_gexit_46_lrlw_joint_nsw.tex
%

\begin{tikzpicture}[scale=0.4,>=stealth]

\definecolor{mycolor1}{rgb}{0,1,1}
\definecolor{mycolor2}{rgb}{1,0,1}

\begin{axis}[
scale only axis,
width=5in,
height=5in,
xmin=0, xmax=1,
ymin=0, ymax=1,
xlabel=\mbox{\Large $\mathsf{h}$},
ylabel=\mbox{\Large $h^{\text{EBP}}$},
legend style={at={(0.41,.99)}},
xmajorgrids,
ymajorgrids]

\addplot [
color=blue,
line width=1.25pt,
solid
]
coordinates{
  (1,0.08)
 (0.926361,0.0973418)
 (0.851165,0.117401)
 (0.789031,0.13861)
 (0.73648,0.160982)
 (0.691803,0.184328)
 (0.653228,0.208569)
 (0.619842,0.233556)
 (0.59024,0.259284)
 (0.564178,0.2855)
 (0.541046,0.312108)
 (0.520477,0.338976)
 (0.501922,0.366062)
 (0.48526,0.393213)
 (0.470306,0.420282)
 (0.456695,0.447301)
 (0.444488,0.474047)
 (0.433502,0.500478)
 (0.423553,0.526513)
 (0.414642,0.552142)
 (0.406769,0.577133)
 (0.399688,0.601618)
 (0.393341,0.625555)
 (0.38797,0.648686)
 (0.383392,0.671074)
 (0.379669,0.692617)
 (0.376739,0.713335)
 (0.374359,0.733364)
 (0.372894,0.752372)
 (0.372101,0.770572)
 (0.372223,0.787735)
 (0.372955,0.804106)
 (0.37442,0.81955)
 (0.376556,0.834148)
 (0.379364,0.84788)
 (0.382904,0.860752)
 (0.387176,0.872744)
 (0.392059,0.883937)
 (0.397552,0.894387)
 (0.403778,0.904032)
 (0.410492,0.913)
 (0.417755,0.921349)
 (0.425567,0.929)
 (0.433929,0.936029)
 (0.442718,0.942574)
 (0.452057,0.948531)
 (0.461761,0.953987)
 (0.471893,0.958982)
 (0.482513,0.963514)
 (0.493438,0.967644)
 (0.504669,0.971437)
 (0.516205,0.974843)
 (0.528106,0.977918)
 (0.540313,0.980703)
 (0.552765,0.983202)
 (0.56546,0.98544)
 (0.578399,0.987442)
 (0.591522,0.989211)
 (0.604889,0.990802)
 (0.618438,0.992186)
 (0.63211,0.993423)
 (0.645965,0.994492)
 (0.659942,0.995429)
 (0.674102,0.99624)
 (0.688324,0.99693)
 (0.702667,0.997526)
 (0.717132,0.998026)
 (0.73172,0.998451)
 (0.746307,0.998798)
 (0.761016,0.999086)
 (0.775726,0.999317)
 (0.790557,0.999503)
 (0.805389,0.999647)
 (0.820281,0.999758)
 (0.835174,0.999841)
 (0.850128,0.999901)
 (0.865081,0.999942)
 (0.880035,0.99997)
 (0.894988,0.999986)
 (0.910003,0.999995)
 (0.925018,0.999999)
 (0.939971,1)
 (0.954986,1)

};
\addlegendentry{\Large JIT - punc $(4,6)$}

\addplot [
color=black,
fill=gray,fill opacity=0.4,
line width=1.25pt,
solid
]
coordinates{
 (0.632368,0)
 (0.632368,0.993443)
 (0.657368,0.995257)
 (0.682368,0.996641)
 (0.707368,0.997689)
 (0.732368,0.998466)
 (0.757368,0.999015)
 (0.782368,0.9994)
 (0.807368,0.999662)
 (0.832368,0.999825)
 (0.857368,0.999921)
 (0.882368,0.999972)
 (0.907368,0.999993)
 (0.932368,0.999999)
 (0.957368,1)
 (1,1)
}|- (axis cs:1,0) -- cycle;
\addlegendentry{\Large MAP - punc $(4,6)$}

\addplot [
color=purple,
solid
]
coordinates{
 (0.632866,0.180582)
 (0.636566,0.206686)
 (0.628326,0.234777)
 (0.631866,0.260683)
 (0.632782,0.286621)
 (0.627349,0.314994)
 (0.639556,0.33756)
 (0.624542,0.369103)
 (0.638519,0.390653)
 (0.624359,0.422437)
 (0.634307,0.445111)
 (0.62796,0.474098)
 (0.630157,0.499918)
 (0.629181,0.526466)
 (0.626617,0.554778)
 (0.638763,0.574832)
 (0.624054,0.609373)
 (0.636749,0.628785)
 (0.623565,0.662924)
 (0.632904,0.683961)
 (0.628814,0.712802)
 (0.629303,0.739339)
 (0.624603,0.768525)
 (0.626129,0.794704)
 (0.63797,0.812502)
 (0.621856,0.8511)
 (0.630401,0.870862)
 (0.621917,0.904499)
 (0.629242,0.925064)
 (0.649078,0.935307)
 (0.666961,0.946907)
 (0.682525,0.960162)
 (0.706207,0.966778)
 (0.730133,0.972784)
 (0.755096,0.977612)
 (0.781036,0.981421)
 (0.808258,0.984196)
 (0.835784,0.986537)
 (0.863738,0.988454)
 (0.892059,0.990055)
 (0.920623,0.99142)
 (0.949371,0.992616)
 (0.978301,0.993691)
 (1,1)
};
\addlegendentry{\Large JIT - $L = 16,w=2$}

\addplot [
color=green,
line width=0.75pt,
solid
]
coordinates{
 (0.632049,0.2)
 (0.632049,0.22681)
 (0.631805,0.259979)
 (0.631744,0.293107)
 (0.631866,0.326206)
 (0.631988,0.359301)
 (0.631439,0.392579)
 (0.631866,0.425575)
 (0.631134,0.458957)
 (0.631073,0.49212)
 (0.631134,0.525257)
 (0.631744,0.558097)
 (0.6315,0.59137)
 (0.631744,0.624368)
 (0.631622,0.657531)
 (0.631866,0.690547)
 (0.631744,0.723746)
 (0.631439,0.757062)
 (0.631073,0.790401)
 (0.629913,0.824365)
 (0.628936,0.858271)
 (0.629974,0.890728)
 (0.640045,0.915981)
 (0.659637,0.933061)
 (0.686004,0.944183)
 (0.714447,0.953099)
 (0.74588,0.959357)
 (0.778289,0.964562)
 (0.811431,0.968873)
 (0.845733,0.972186)
 (0.880706,0.974908)
 (0.915862,0.977361)
 (0.951263,0.979537)
 (0.986846,0.981544)
 (1,0.9827)
};
\addlegendentry{\Large JIT - $L = 32,w=4$}

\addplot [
color=blue,
line width=1.25pt,
dash pattern=on 1pt off 3pt on 3pt off 3pt
]
coordinates{
 (0.372101,0)
 (0.372101,0.770572)

};

\end{axis}

\end{tikzpicture}


%% file: acpr_36_spatial_mac.tex
\begin{tikzpicture}[scale=0.37]

\begin{axis}[
scale only axis,
width=5in,
height=5in,
xmin=0.6, xmax=2.4,
ymin=0.6, ymax=2.4,
xmajorgrids,
ymajorgrids,
xlabel={\LARGE$h_1$},
ylabel={\LARGE$h_2$},
legend style = {at={(0.54,0.16)}}
]

\addplot [
color=red,
solid
]
coordinates{
 (1.03,1.45)
 (1.04,1.45)
 (1.05,1.44)
 (1.06,1.43)
 (1.07,1.42)
 (1.08,1.42)
 (1.09,1.41)
 (1.1,1.4)
 (1.11,1.39)
 (1.12,1.39)
 (1.13,1.38)
 (1.14,1.37)
 (1.15,1.36)
 (1.16,1.35)
 (1.17,1.35)
 (1.18,1.34)
 (1.19,1.33)
 (1.2,1.32)
 (1.21,1.31)
 (1.22,1.3)
 (1.23,1.29)
 (1.24,1.28)
 (1.25,1.27)
 (1.26,1.26)
 (1.27,1.25)
 (1.28,1.24)
 (1.29,1.23)
 (1.3,1.22)
 (1.31,1.21)
 (1.32,1.2)
 (1.33,1.19)
 (1.34,1.18)
 (1.35,1.17)
 (1.35,1.16)
 (1.36,1.15)
 (1.37,1.14)
 (1.38,1.13)
 (1.39,1.12)
 (1.39,1.11)
 (1.4,1.1)
 (1.41,1.09)
 (1.42,1.08)
 (1.42,1.07)
 (1.43,1.06)
 (1.44,1.05)
 (1.45,1.04)
 (1.45,1.03)
 (1.46,1.03)
 (1.47,1.03)
 (1.48,1.03)
 (1.49,1.03)
 (1.5,1.03)
 (1.51,1.03)
 (1.52,1.03)
 (1.53,1.03)
 (1.54,1.03)
 (1.55,1.03)
 (1.56,1.03)
 (1.57,1.03)
 (1.58,1.03)
 (1.59,1.03)
 (1.6,1.03)
 (1.61,1.03)
 (1.62,1.03)
 (1.63,1.03)
 (1.64,1.03)
 (1.65,1.03)
 (1.66,1.03)
 (1.67,1.03)
 (1.68,1.03)
 (1.69,1.03)
 (1.7,1.03)
 (1.71,1.03)
 (1.72,1.03)
 (1.73,1.03)
 (1.74,1.03)
 (1.75,1.03)
 (1.76,1.03)
 (1.77,1.03)
 (1.78,1.03)
 (1.79,1.03)
 (1.8,1.03)
 (1.81,1.03)
 (1.82,1.03)
 (1.83,1.03)
 (1.84,1.03)
 (1.85,1.03)
 (1.86,1.03)
 (1.87,1.03)
 (1.88,1.03)
 (1.89,1.03)
 (1.9,1.03)
 (1.91,1.03)
 (1.92,1.03)
 (1.93,1.03)
 (1.94,1.03)
 (1.95,1.03)
 (1.96,1.03)
 (1.97,1.03)
 (1.98,1.03)
 (1.99,1.03)
 (2,1.03)
 (2.01,1.03)
 (2.02,1.03)
 (2.03,1.03)
 (2.04,1.03)
 (2.05,1.03)
 (2.06,1.03)
 (2.07,1.03)
 (2.08,1.03)
 (2.09,1.03)
 (2.1,1.03)
 (2.11,1.03)
 (2.12,1.03)
 (2.13,1.03)
 (2.14,1.03)
 (2.15,1.03)
 (2.16,1.03)
 (2.17,1.03)
 (2.18,1.03)
 (2.19,1.03)
 (2.2,1.03)
 (2.21,1.03)
 (2.22,1.03)
 (2.23,1.03)
 (2.24,1.03)
 (2.25,1.03)
 (2.26,1.03)
 (2.27,1.03)
 (2.28,1.03)
 (2.29,1.03)
 (2.3,1.03)
 (2.31,1.03)
 (2.32,1.03)
 (2.33,1.03)
 (2.34,1.03)
 (2.35,1.03)
 (2.36,1.03)
 (2.37,1.03)
 (2.38,1.03)
 (2.39,1.03)
 (2.4,1.03)
 (2.41,1.03)
 (2.42,1.03)
 (2.43,1.03)
 (2.44,1.03)
 (2.45,1.03)
 (2.46,1.03)
 (2.47,1.03)
 (2.48,1.03)
 (2.49,1.03)
 (2.5,1.03)
 (2.5,2.5)
 (1.03,2.5)
 (1.03,2.49)
 (1.03,2.48)
 (1.03,2.47)
 (1.03,2.46)
 (1.03,2.45)
 (1.03,2.44)
 (1.03,2.43)
 (1.03,2.42)
 (1.03,2.41)
 (1.03,2.4)
 (1.03,2.39)
 (1.03,2.38)
 (1.03,2.37)
 (1.03,2.36)
 (1.03,2.35)
 (1.03,2.34)
 (1.03,2.33)
 (1.03,2.32)
 (1.03,2.31)
 (1.03,2.3)
 (1.03,2.29)
 (1.03,2.28)
 (1.03,2.27)
 (1.03,2.26)
 (1.03,2.25)
 (1.03,2.24)
 (1.03,2.23)
 (1.03,2.22)
 (1.03,2.21)
 (1.03,2.2)
 (1.03,2.19)
 (1.03,2.18)
 (1.03,2.17)
 (1.03,2.16)
 (1.03,2.15)
 (1.03,2.14)
 (1.03,2.13)
 (1.03,2.12)
 (1.03,2.11)
 (1.03,2.1)
 (1.03,2.09)
 (1.03,2.08)
 (1.03,2.07)
 (1.03,2.06)
 (1.03,2.05)
 (1.03,2.04)
 (1.03,2.03)
 (1.03,2.02)
 (1.03,2.01)
 (1.03,2)
 (1.03,1.99)
 (1.03,1.98)
 (1.03,1.97)
 (1.03,1.96)
 (1.03,1.95)
 (1.03,1.94)
 (1.03,1.93)
 (1.03,1.92)
 (1.03,1.91)
 (1.03,1.9)
 (1.03,1.89)
 (1.03,1.88)
 (1.03,1.87)
 (1.03,1.86)
 (1.03,1.85)
 (1.03,1.84)
 (1.03,1.83)
 (1.03,1.82)
 (1.03,1.81)
 (1.03,1.8)
 (1.03,1.79)
 (1.03,1.78)
 (1.03,1.77)
 (1.03,1.76)
 (1.03,1.75)
 (1.03,1.74)
 (1.03,1.73)
 (1.03,1.72)
 (1.03,1.71)
 (1.03,1.7)
 (1.03,1.69)
 (1.03,1.68)
 (1.03,1.67)
 (1.03,1.66)
 (1.03,1.65)
 (1.03,1.64)
 (1.03,1.63)
 (1.03,1.62)
 (1.03,1.61)
 (1.03,1.6)
 (1.03,1.59)
 (1.03,1.58)
 (1.03,1.57)
 (1.03,1.56)
 (1.03,1.55)
 (1.03,1.54)
 (1.03,1.53)
 (1.03,1.52)
 (1.03,1.51)
 (1.03,1.5)
 (1.03,1.49)
 (1.03,1.48)
 (1.03,1.47)
 (1.03,1.46)
 (1.03,1.45)

};
\addlegendentry{\Large Capacity Region for rate $0.5$}

\addplot [
black,fill=gray,fill opacity=0.2
]
coordinates{
 (1.13599,2.5)
 (1.13599,2.0107)
 (1.13599,1.99934)
 (1.13599,1.98798)
 (1.13599,1.97662)
 (1.13599,1.96526)
 (1.13599,1.9539)
 (1.13599,1.94254)
 (1.13599,1.93118)
 (1.13599,1.91982)
 (1.13599,1.90846)
 (1.13599,1.8971)
 (1.13599,1.88574)
 (1.13599,1.87438)
 (1.13599,1.86302)
 (1.13599,1.85166)
 (1.13599,1.8403)
 (1.13599,1.82894)
 (1.13599,1.81758)
 (1.13599,1.80622)
 (1.13599,1.79486)
 (1.13599,1.7835)
 (1.13599,1.77214)
 (1.13599,1.76078)
 (1.13599,1.74942)
 (1.13599,1.73806)
 (1.13599,1.7267)
 (1.13599,1.71534)
 (1.13599,1.70398)
 (1.13599,1.69262)
 (1.13599,1.68126)
 (1.13599,1.6699)
 (1.13599,1.65854)
 (1.13599,1.64718)
 (1.13599,1.63582)
 (1.14551,1.63808)
 (1.15723,1.64326)
 (1.16895,1.64821)
 (1.17773,1.64883)
 (1.18945,1.65334)
 (1.20117,1.65762)
 (1.21289,1.66166)
 (1.22461,1.66547)
 (1.23926,1.673)
 (1.25098,1.67631)
 (1.2627,1.67938)
 (1.27441,1.68223)
 (1.28613,1.68483)
 (1.29932,1.68911)
 (1.31104,1.69124)
 (1.32385,1.69453)
 (1.3418,1.70408)
 (1.35352,1.70543)
 (1.36523,1.70654)
 (1.37695,1.70742)
 (1.3916,1.71167)
 (1.40479,1.71384)
 (1.4176,1.7153)
 (1.43555,1.72266)
 (1.44727,1.72225)
 (1.46191,1.72506)
 (1.47363,1.72415)
 (1.48755,1.72556)
 (1.50586,1.73174)
 (1.51758,1.73004)
 (1.53223,1.73142)
 (1.54395,1.72922)
 (1.55786,1.72923)
 (1.57617,1.73379)
 (1.58789,1.7308)
 (1.59961,1.72758)
 (1.61133,1.72412)
 (1.62598,1.72354)
 (1.6377,1.71958)
 (1.64941,1.71539)
 (1.66113,1.71097)
 (1.66992,1.70332)
 (1.68164,1.69846)
 (1.68732,1.68732)
 (1.68732,1.68732)
 (1.69846,1.68164)
 (1.70332,1.66992)
 (1.71097,1.66113)
 (1.71539,1.64941)
 (1.71958,1.6377)
 (1.72354,1.62598)
 (1.72412,1.61133)
 (1.72758,1.59961)
 (1.7308,1.58789)
 (1.73379,1.57617)
 (1.72923,1.55786)
 (1.72922,1.54395)
 (1.73142,1.53223)
 (1.73004,1.51758)
 (1.73174,1.50586)
 (1.72556,1.48755)
 (1.72415,1.47363)
 (1.72506,1.46191)
 (1.72225,1.44727)
 (1.72266,1.43555)
 (1.7153,1.4176)
 (1.71384,1.40479)
 (1.71167,1.3916)
 (1.70742,1.37695)
 (1.70654,1.36523)
 (1.70543,1.35352)
 (1.70408,1.3418)
 (1.69453,1.32385)
 (1.69124,1.31104)
 (1.68911,1.29932)
 (1.68483,1.28613)
 (1.68223,1.27441)
 (1.67938,1.2627)
 (1.67631,1.25098)
 (1.673,1.23926)
 (1.66547,1.22461)
 (1.66166,1.21289)
 (1.65762,1.20117)
 (1.65334,1.18945)
 (1.64883,1.17773)
 (1.64821,1.16895)
 (1.64326,1.15723)
 (1.63808,1.14551)
 (1.63582,1.13599)
 (1.64718,1.13599)
 (1.65854,1.13599)
 (1.6699,1.13599)
 (1.68126,1.13599)
 (1.69262,1.13599)
 (1.70398,1.13599)
 (1.71534,1.13599)
 (1.7267,1.13599)
 (1.73806,1.13599)
 (1.74942,1.13599)
 (1.76078,1.13599)
 (1.77214,1.13599)
 (1.7835,1.13599)
 (1.79486,1.13599)
 (1.80622,1.13599)
 (1.81758,1.13599)
 (1.82894,1.13599)
 (1.8403,1.13599)
 (1.85166,1.13599)
 (1.86302,1.13599)
 (1.87438,1.13599)
 (1.88574,1.13599)
 (1.8971,1.13599)
 (1.90846,1.13599)
 (1.91982,1.13599)
 (1.93118,1.13599)
 (1.94254,1.13599)
 (1.9539,1.13599)
 (1.96526,1.13599)
 (1.97662,1.13599)
 (1.98798,1.13599)
 (1.99934,1.13599)
 (2.0107,1.13599)
 (2.5,1.13599)

}|- (axis cs:2.5,2.5) -- cycle;
\addlegendentry{\Large ACPR (DE) - $(3,6)$}

\addplot [
black,fill=gray,fill opacity=0.4
]
coordinates{
 (1.06055,2.5)
 (1.06055,2.12109)
 (1.06055,1.46851)
 (1.07031,1.46062)
 (1.08594,1.46023)
 (1.08984,1.44368)
 (1.10156,1.43717)
 (1.10889,1.42455)
 (1.12109,1.41781)
 (1.13281,1.40998)
 (1.14056,1.39681)
 (1.15234,1.38819)
 (1.16406,1.37903)
 (1.17969,1.37394)
 (1.18652,1.35818)
 (1.19922,1.34872)
 (1.21094,1.33768)
 (1.22656,1.33041)
 (1.23389,1.31368)
 (1.24609,1.30175)
 (1.26172,1.29284)
 (1.27344,1.27938)
 (1.27938,1.27344)
 (1.29284,1.26172)
 (1.30175,1.24609)
 (1.31368,1.23389)
 (1.33041,1.22656)
 (1.33768,1.21094)
 (1.34872,1.19922)
 (1.35818,1.18652)
 (1.37394,1.17969)
 (1.37903,1.16406)
 (1.38819,1.15234)
 (1.39681,1.14056)
 (1.40998,1.13281)
 (1.41781,1.12109)
 (1.42455,1.10889)
 (1.43717,1.10156)
 (1.44368,1.08984)
 (1.46023,1.08594)
 (1.46062,1.07031)
 (1.46851,1.06055)
 (2.12109,1.06055)
(2.5,1.06055)
}|- (axis cs:2.5,2.5) -- cycle;
\addlegendentry{\Large ACPR (DE) - $(3,6,32,4)$}

\end{axis}
\end{tikzpicture}

%% file: 46_lrlw_acpr.tex
%

\begin{tikzpicture}[scale=0.37]

\begin{axis}[
scale only axis,
width=5in,
height=5in,
xmin=0, xmax=1,
ymin=0, ymax=1,
xlabel={\LARGE$\epsilon_1$},
ylabel={\LARGE$\epsilon_2$},
xmajorgrids,
ymajorgrids,
legend style = {at={(0.98,0.98)}}
]

\addplot [
color=red,
solid,
line width=1.2pt
]
coordinates{
 (0,0.760099)
 (0.13005,0.760099)
 (0.260099,0.760099)
 (0.390149,0.760099)
 (0.520199,0.760099)
 (0.520199,0.760099)
 (0.580174,0.700124)
 (0.640149,0.640149)
 (0.700124,0.580174)
 (0.760099,0.520199)
 (0.760099,0)
 (0.760099,0.13005)
 (0.760099,0.260099)
 (0.760099,0.390149)
 (0.760099,0.520199)

};
\addlegendentry{\Large SW Region for rate $0.4798$}

\addplot [
color=blue,
solid,
line width=1.2pt
]
coordinates{
 (0,0.75)
 (0.125,0.75)
 (0.25,0.75)
 (0.375,0.75)
 (0.5,0.75)
 (0.5,0.75)
 (0.5625,0.6875)
 (0.625,0.625)
 (0.6875,0.5625)
 (0.75,0.5)
 (0.75,0)
 (0.75,0.125)
 (0.75,0.25)
 (0.75,0.375)
 (0.75,0.5)

};
\addlegendentry{\Large SW Region for rate $0.5$}

\addplot [
black,fill=gray,fill opacity=0.2
]
coordinates{
 (0.743434,0)
 (0.743434,0.49697)
 (0.739394,0.50101)
 (0.739394,0.505051)
 (0.735354,0.509091)
 (0.731313,0.513131)
 (0.727273,0.517172)
 (0.723232,0.521212)
 (0.719192,0.525253)
 (0.715152,0.529293)
 (0.711111,0.533333)
 (0.707071,0.537374)
 (0.70303,0.541414)
 (0.69899,0.545455)
 (0.694949,0.549495)
 (0.690909,0.553535)
 (0.686869,0.557576)
 (0.682828,0.561616)
 (0.678788,0.565657)
 (0.674747,0.569697)
 (0.670707,0.573737)
 (0.666667,0.577778)
 (0.662626,0.581818)
 (0.658586,0.585859)
 (0.654545,0.585859)
 (0.650505,0.589899)
 (0.646465,0.593939)
 (0.642424,0.59798)
 (0.638384,0.60202)
 (0.634343,0.606061)
 (0.630303,0.610101)
 (0.626263,0.614141)
 (0.622222,0.618182)
 (0.618182,0.622222)
 (0.614141,0.626263)
 (0.610101,0.630303)
 (0.606061,0.634343)
 (0.60202,0.638384)
 (0.59798,0.642424)
 (0.593939,0.646465)
 (0.589899,0.650505)
 (0.585859,0.654545)
 (0.585859,0.658586)
 (0.581818,0.662626)
 (0.577778,0.666667)
 (0.573737,0.670707)
 (0.569697,0.674747)
 (0.565657,0.678788)
 (0.561616,0.682828)
 (0.557576,0.686869)
 (0.553535,0.690909)
 (0.549495,0.694949)
 (0.545455,0.69899)
 (0.541414,0.70303)
 (0.537374,0.707071)
 (0.533333,0.711111)
 (0.529293,0.715152)
 (0.525253,0.719192)
 (0.521212,0.723232)
 (0.517172,0.727273)
 (0.513131,0.731313)
 (0.509091,0.735354)
 (0.505051,0.739394)
 (0.50101,0.739394)
 (0.49697,0.743434)
 (0.492929,0.743434)
 (0.488889,0.743434)
 (0.484848,0.743434)
 (0.480808,0.743434)
 (0.476768,0.743434)
 (0.472727,0.743434)
 (0.468687,0.743434)
 (0.464646,0.743434)
 (0.460606,0.743434)
 (0.456566,0.743434)
 (0.452525,0.743434)
 (0.448485,0.743434)
 (0.444444,0.743434)
 (0.440404,0.743434)
 (0.436364,0.743434)
 (0.432323,0.743434)
 (0.428283,0.743434)
 (0.424242,0.743434)
 (0.420202,0.743434)
 (0.416162,0.743434)
 (0.412121,0.743434)
 (0.408081,0.743434)
 (0.40404,0.743434)
 (0.4,0.743434)
 (0,0.743434)
}|- (axis cs:0,0) -- cycle;
\addlegendentry{\Large ACPR (DE) - $(4,6,64,10)$}

\addplot [
black,fill=gray,fill opacity=0.4
]
coordinates{
 (0.5,0)
 (0.5,0.01)
 (0.5,0.02)
 (0.5,0.03)
 (0.5,0.04)
 (0.5,0.05)
 (0.5,0.06)
 (0.5,0.07)
 (0.5,0.08)
 (0.5,0.09)
 (0.5,0.1)
 (0.5,0.11)
 (0.5,0.12)
 (0.5,0.13)
 (0.5,0.14)
 (0.5,0.15)
 (0.5,0.16)
 (0.5,0.17)
 (0.5,0.18)
 (0.5,0.19)
 (0.5,0.2)
 (0.5,0.21)
 (0.5,0.22)
 (0.5,0.23)
 (0.5,0.24)
 (0.5,0.25)
 (0.5,0.26)
 (0.5,0.27)
 (0.49,0.27)
 (0.48,0.27)
 (0.47,0.27)
 (0.46,0.27)
 (0.45,0.27)
 (0.44,0.28)
 (0.43,0.28)
 (0.42,0.28)
 (0.41,0.28)
 (0.4,0.29)
 (0.39,0.29)
 (0.38,0.29)
 (0.37,0.3)
 (0.36,0.3)
 (0.35,0.31)
 (0.34,0.32)
 (0.33,0.33)
 (0.32,0.34)
 (0.31,0.35)
 (0.3,0.36)
 (0.3,0.37)
 (0.29,0.38)
 (0.29,0.39)
 (0.29,0.4)
 (0.28,0.41)
 (0.28,0.42)
 (0.28,0.43)
 (0.28,0.44)
 (0.27,0.45)
 (0.27,0.46)
 (0.27,0.47)
 (0.27,0.48)
 (0.27,0.49)
 (0.27,0.5)
 (0.26,0.5)
 (0.25,0.5)
 (0.24,0.5)
 (0.23,0.5)
 (0.22,0.5)
 (0.21,0.5)
 (0.2,0.5)
 (0.19,0.5)
 (0.18,0.5)
 (0.17,0.5)
 (0.16,0.5)
 (0.15,0.5)
 (0.14,0.5)
 (0.13,0.5)
 (0.12,0.5)
 (0.11,0.5)
 (0.1,0.5)
 (0.09,0.5)
 (0.08,0.5)
 (0.07,0.5)
 (0.06,0.5)
 (0.05,0.5)
 (0.04,0.5)
 (0.03,0.5)
 (0.02,0.5)
 (0.01,0.5)
 (0,0.5)

}|- (axis cs:0,0) -- cycle;
\addlegendentry{\Large ACPR (DE) - (4,6)};

\end{axis}

\end{tikzpicture}


%% file: 46_lrlw_acpr_awgn.tex
\begin{tikzpicture}[scale=0.37]

\begin{axis}[
scale only axis,
width=5in,
height=5in,
xmin=-5, xmax=3,
ymin=-5, ymax=3,
xlabel={\LARGE$\text{SNR}_1$},
ylabel={\LARGE$\text{SNR}_2$},
xmajorgrids,
ymajorgrids,
legend style = {at={(0.52,0.98)}}
]

\addplot [
color=red,
solid,
]
coordinates{
 (-4.2,3)
 (-4.2,0.1)
 (-4.19,0.1)
 (-4.18,0.09)
 (-4.17,0.09)
 (-4.16,0.08)
 (-4.15,0.07)
 (-4.14,0.07)
 (-4.13,0.06)
 (-4.12,0.05)
 (-4.11,0.05)
 (-4.1,0.04)
 (-4.09,0.04)
 (-4.08,0.03)
 (-4.07,0.02)
 (-4.06,0.02)
 (-4.05,0.01)
 (-4.04,0.01)
 (-4.03,0)
 (-4.02,-0.01)
 (-4.01,-0.01)
 (-4,-0.02)
 (-3.99,-0.03)
 (-3.98,-0.03)
 (-3.97,-0.04)
 (-3.96,-0.04)
 (-3.95,-0.05)
 (-3.94,-0.06)
 (-3.93,-0.06)
 (-3.92,-0.07)
 (-3.91,-0.08)
 (-3.9,-0.08)
 (-3.89,-0.09)
 (-3.88,-0.09)
 (-3.87,-0.1)
 (-3.86,-0.11)
 (-3.85,-0.11)
 (-3.84,-0.12)
 (-3.83,-0.13)
 (-3.82,-0.13)
 (-3.81,-0.14)
 (-3.8,-0.15)
 (-3.79,-0.15)
 (-3.78,-0.16)
 (-3.77,-0.17)
 (-3.76,-0.17)
 (-3.75,-0.18)
 (-3.74,-0.19)
 (-3.73,-0.19)
 (-3.72,-0.2)
 (-3.71,-0.21)
 (-3.7,-0.21)
 (-3.69,-0.22)
 (-3.68,-0.23)
 (-3.67,-0.23)
 (-3.66,-0.24)
 (-3.65,-0.25)
 (-3.64,-0.25)
 (-3.63,-0.26)
 (-3.62,-0.27)
 (-3.61,-0.27)
 (-3.6,-0.28)
 (-3.59,-0.29)
 (-3.58,-0.29)
 (-3.57,-0.3)
 (-3.56,-0.31)
 (-3.55,-0.31)
 (-3.54,-0.32)
 (-3.53,-0.33)
 (-3.52,-0.33)
 (-3.51,-0.34)
 (-3.5,-0.35)
 (-3.49,-0.35)
 (-3.48,-0.36)
 (-3.47,-0.37)
 (-3.46,-0.38)
 (-3.45,-0.38)
 (-3.44,-0.39)
 (-3.43,-0.4)
 (-3.42,-0.4)
 (-3.41,-0.41)
 (-3.4,-0.42)
 (-3.39,-0.42)
 (-3.38,-0.43)
 (-3.37,-0.44)
 (-3.36,-0.45)
 (-3.35,-0.45)
 (-3.34,-0.46)
 (-3.33,-0.47)
 (-3.32,-0.47)
 (-3.31,-0.48)
 (-3.3,-0.49)
 (-3.29,-0.5)
 (-3.28,-0.5)
 (-3.27,-0.51)
 (-3.26,-0.52)
 (-3.25,-0.52)
 (-3.24,-0.53)
 (-3.23,-0.54)
 (-3.22,-0.55)
 (-3.21,-0.55)
 (-3.2,-0.56)
 (-3.19,-0.57)
 (-3.18,-0.58)
 (-3.17,-0.58)
 (-3.16,-0.59)
 (-3.15,-0.6)
 (-3.14,-0.61)
 (-3.13,-0.61)
 (-3.12,-0.62)
 (-3.11,-0.63)
 (-3.1,-0.64)
 (-3.09,-0.64)
 (-3.08,-0.65)
 (-3.07,-0.66)
 (-3.06,-0.67)
 (-3.05,-0.67)
 (-3.04,-0.68)
 (-3.03,-0.69)
 (-3.02,-0.7)
 (-3.01,-0.7)
 (-3,-0.71)
 (-2.99,-0.72)
 (-2.98,-0.73)
 (-2.97,-0.73)
 (-2.96,-0.74)
 (-2.95,-0.75)
 (-2.94,-0.76)
 (-2.93,-0.77)
 (-2.92,-0.77)
 (-2.91,-0.78)
 (-2.9,-0.79)
 (-2.89,-0.8)
 (-2.88,-0.8)
 (-2.87,-0.81)
 (-2.86,-0.82)
 (-2.85,-0.83)
 (-2.84,-0.84)
 (-2.83,-0.84)
 (-2.82,-0.85)
 (-2.81,-0.86)
 (-2.8,-0.87)
 (-2.79,-0.88)
 (-2.78,-0.88)
 (-2.77,-0.89)
 (-2.76,-0.9)
 (-2.75,-0.91)
 (-2.74,-0.92)
 (-2.73,-0.92)
 (-2.72,-0.93)
 (-2.71,-0.94)
 (-2.7,-0.95)
 (-2.69,-0.96)
 (-2.68,-0.97)
 (-2.67,-0.97)
 (-2.66,-0.98)
 (-2.65,-0.99)
 (-2.64,-1)
 (-2.63,-1.01)
 (-2.62,-1.01)
 (-2.61,-1.02)
 (-2.6,-1.03)
 (-2.59,-1.04)
 (-2.58,-1.05)
 (-2.57,-1.06)
 (-2.56,-1.06)
 (-2.55,-1.07)
 (-2.54,-1.08)
 (-2.53,-1.09)
 (-2.52,-1.1)
 (-2.51,-1.11)
 (-2.5,-1.12)
 (-2.49,-1.12)
 (-2.48,-1.13)
 (-2.47,-1.14)
 (-2.46,-1.15)
 (-2.45,-1.16)
 (-2.44,-1.17)
 (-2.43,-1.18)
 (-2.42,-1.18)
 (-2.41,-1.19)
 (-2.4,-1.2)
 (-2.39,-1.21)
 (-2.38,-1.22)
 (-2.37,-1.23)
 (-2.36,-1.24)
 (-2.35,-1.25)
 (-2.34,-1.25)
 (-2.33,-1.26)
 (-2.32,-1.27)
 (-2.31,-1.28)
 (-2.3,-1.29)
 (-2.29,-1.3)
 (-2.28,-1.31)
 (-2.27,-1.32)
 (-2.26,-1.33)
 (-2.25,-1.33)
 (-2.24,-1.34)
 (-2.23,-1.35)
 (-2.22,-1.36)
 (-2.21,-1.37)
 (-2.2,-1.38)
 (-2.19,-1.39)
 (-2.18,-1.4)
 (-2.17,-1.41)
 (-2.16,-1.42)
 (-2.15,-1.43)
 (-2.14,-1.43)
 (-2.13,-1.44)
 (-2.12,-1.45)
 (-2.11,-1.46)
 (-2.1,-1.47)
 (-2.09,-1.48)
 (-2.08,-1.49)
 (-2.07,-1.5)
 (-2.06,-1.51)
 (-2.05,-1.52)
 (-2.04,-1.53)
 (-2.03,-1.54)
 (-2.02,-1.55)
 (-2.01,-1.56)
 (-2,-1.57)
 (-1.99,-1.58)
 (-1.98,-1.58)
 (-1.97,-1.59)
 (-1.96,-1.6)
 (-1.95,-1.61)
 (-1.94,-1.62)
 (-1.93,-1.63)
 (-1.92,-1.64)
 (-1.91,-1.65)
 (-1.9,-1.66)
 (-1.89,-1.67)
 (-1.88,-1.68)
 (-1.87,-1.69)
 (-1.86,-1.7)
 (-1.85,-1.71)
 (-1.84,-1.72)
 (-1.83,-1.73)
 (-1.82,-1.74)
 (-1.81,-1.75)
 (-1.8,-1.76)
 (-1.79,-1.77)
 (-1.78,-1.78)
 (-1.77,-1.79)
 (-1.76,-1.8)
 (-1.75,-1.81)
 (-1.74,-1.82)
 (-1.73,-1.83)
 (-1.72,-1.84)
 (-1.71,-1.85)
 (-1.7,-1.86)
 (-1.69,-1.87)
 (-1.68,-1.88)
 (-1.67,-1.89)
 (-1.66,-1.9)
 (-1.65,-1.91)
 (-1.64,-1.92)
 (-1.63,-1.93)
 (-1.62,-1.94)
 (-1.61,-1.95)
 (-1.6,-1.96)
 (-1.59,-1.97)
 (-1.58,-1.98)
 (-1.58,-1.99)
 (-1.57,-2)
 (-1.56,-2.01)
 (-1.55,-2.02)
 (-1.54,-2.03)
 (-1.53,-2.04)
 (-1.52,-2.05)
 (-1.51,-2.06)
 (-1.5,-2.07)
 (-1.49,-2.08)
 (-1.48,-2.09)
 (-1.47,-2.1)
 (-1.46,-2.11)
 (-1.45,-2.12)
 (-1.44,-2.13)
 (-1.43,-2.14)
 (-1.43,-2.15)
 (-1.42,-2.16)
 (-1.41,-2.17)
 (-1.4,-2.18)
 (-1.39,-2.19)
 (-1.38,-2.2)
 (-1.37,-2.21)
 (-1.36,-2.22)
 (-1.35,-2.23)
 (-1.34,-2.24)
 (-1.33,-2.25)
 (-1.33,-2.26)
 (-1.32,-2.27)
 (-1.31,-2.28)
 (-1.3,-2.29)
 (-1.29,-2.3)
 (-1.28,-2.31)
 (-1.27,-2.32)
 (-1.26,-2.33)
 (-1.25,-2.34)
 (-1.25,-2.35)
 (-1.24,-2.36)
 (-1.23,-2.37)
 (-1.22,-2.38)
 (-1.21,-2.39)
 (-1.2,-2.4)
 (-1.19,-2.41)
 (-1.18,-2.42)
 (-1.18,-2.43)
 (-1.17,-2.44)
 (-1.16,-2.45)
 (-1.15,-2.46)
 (-1.14,-2.47)
 (-1.13,-2.48)
 (-1.12,-2.49)
 (-1.12,-2.5)
 (-1.11,-2.51)
 (-1.1,-2.52)
 (-1.09,-2.53)
 (-1.08,-2.54)
 (-1.07,-2.55)
 (-1.06,-2.56)
 (-1.06,-2.57)
 (-1.05,-2.58)
 (-1.04,-2.59)
 (-1.03,-2.6)
 (-1.02,-2.61)
 (-1.01,-2.62)
 (-1.01,-2.63)
 (-1,-2.64)
 (-0.99,-2.65)
 (-0.98,-2.66)
 (-0.97,-2.67)
 (-0.97,-2.68)
 (-0.96,-2.69)
 (-0.95,-2.7)
 (-0.94,-2.71)
 (-0.93,-2.72)
 (-0.92,-2.73)
 (-0.92,-2.74)
 (-0.91,-2.75)
 (-0.9,-2.76)
 (-0.89,-2.77)
 (-0.88,-2.78)
 (-0.88,-2.79)
 (-0.87,-2.8)
 (-0.86,-2.81)
 (-0.85,-2.82)
 (-0.84,-2.83)
 (-0.84,-2.84)
 (-0.83,-2.85)
 (-0.82,-2.86)
 (-0.81,-2.87)
 (-0.8,-2.88)
 (-0.8,-2.89)
 (-0.79,-2.9)
 (-0.78,-2.91)
 (-0.77,-2.92)
 (-0.77,-2.93)
 (-0.76,-2.94)
 (-0.75,-2.95)
 (-0.74,-2.96)
 (-0.73,-2.97)
 (-0.73,-2.98)
 (-0.72,-2.99)
 (-0.71,-3)
 (-0.7,-3.01)
 (-0.7,-3.02)
 (-0.69,-3.03)
 (-0.68,-3.04)
 (-0.67,-3.05)
 (-0.67,-3.06)
 (-0.66,-3.07)
 (-0.65,-3.08)
 (-0.64,-3.09)
 (-0.64,-3.1)
 (-0.63,-3.11)
 (-0.62,-3.12)
 (-0.61,-3.13)
 (-0.61,-3.14)
 (-0.6,-3.15)
 (-0.59,-3.16)
 (-0.58,-3.17)
 (-0.58,-3.18)
 (-0.57,-3.19)
 (-0.56,-3.2)
 (-0.55,-3.21)
 (-0.55,-3.22)
 (-0.54,-3.23)
 (-0.53,-3.24)
 (-0.52,-3.25)
 (-0.52,-3.26)
 (-0.51,-3.27)
 (-0.5,-3.28)
 (-0.5,-3.29)
 (-0.49,-3.3)
 (-0.48,-3.31)
 (-0.47,-3.32)
 (-0.47,-3.33)
 (-0.46,-3.34)
 (-0.45,-3.35)
 (-0.45,-3.36)
 (-0.44,-3.37)
 (-0.43,-3.38)
 (-0.42,-3.39)
 (-0.42,-3.4)
 (-0.41,-3.41)
 (-0.4,-3.42)
 (-0.4,-3.43)
 (-0.39,-3.44)
 (-0.38,-3.45)
 (-0.38,-3.46)
 (-0.37,-3.47)
 (-0.36,-3.48)
 (-0.35,-3.49)
 (-0.35,-3.5)
 (-0.34,-3.51)
 (-0.33,-3.52)
 (-0.33,-3.53)
 (-0.32,-3.54)
 (-0.31,-3.55)
 (-0.31,-3.56)
 (-0.3,-3.57)
 (-0.29,-3.58)
 (-0.29,-3.59)
 (-0.28,-3.6)
 (-0.27,-3.61)
 (-0.27,-3.62)
 (-0.26,-3.63)
 (-0.25,-3.64)
 (-0.25,-3.65)
 (-0.24,-3.66)
 (-0.23,-3.67)
 (-0.23,-3.68)
 (-0.22,-3.69)
 (-0.21,-3.7)
 (-0.21,-3.71)
 (-0.2,-3.72)
 (-0.19,-3.73)
 (-0.19,-3.74)
 (-0.18,-3.75)
 (-0.17,-3.76)
 (-0.17,-3.77)
 (-0.16,-3.78)
 (-0.15,-3.79)
 (-0.15,-3.8)
 (-0.14,-3.81)
 (-0.13,-3.82)
 (-0.13,-3.83)
 (-0.12,-3.84)
 (-0.11,-3.85)
 (-0.11,-3.86)
 (-0.1,-3.87)
 (-0.09,-3.88)
 (-0.09,-3.89)
 (-0.08,-3.9)
 (-0.08,-3.91)
 (-0.07,-3.92)
 (-0.06,-3.93)
 (-0.06,-3.94)
 (-0.05,-3.95)
 (-0.04,-3.96)
 (-0.04,-3.97)
 (-0.03,-3.98)
 (-0.03,-3.99)
 (-0.02,-4)
 (-0.01,-4.01)
 (-0.01,-4.02)
 (0,-4.03)
 (0.01,-4.04)
 (0.01,-4.05)
 (0.02,-4.06)
 (0.02,-4.07)
 (0.03,-4.08)
 (0.04,-4.09)
 (0.04,-4.1)
 (0.05,-4.11)
 (0.05,-4.12)
 (0.06,-4.13)
 (0.07,-4.14)
 (0.07,-4.15)
 (0.08,-4.16)
 (0.09,-4.17)
 (0.09,-4.18)
 (0.1,-4.19)
 (0.1,-4.2)
 (0.11,-4.2)
 (3,-4.2)

};
\addlegendentry{\Large SW Region for rate $0.4934$}

\addplot [
color=blue,
line width=0.2pt,
solid
]
coordinates{
 (-4.13,3)
 (-4.13,0.19)
 (-4.12,0.18)
 (-4.11,0.18)
 (-4.1,0.17)
 (-4.09,0.16)
 (-4.08,0.16)
 (-4.07,0.15)
 (-4.06,0.15)
 (-4.05,0.14)
 (-4.04,0.13)
 (-4.03,0.13)
 (-4.02,0.12)
 (-4.01,0.12)
 (-4,0.11)
 (-3.99,0.1)
 (-3.98,0.1)
 (-3.97,0.09)
 (-3.96,0.08)
 (-3.95,0.08)
 (-3.94,0.07)
 (-3.93,0.07)
 (-3.92,0.06)
 (-3.91,0.05)
 (-3.9,0.05)
 (-3.89,0.04)
 (-3.88,0.03)
 (-3.87,0.03)
 (-3.86,0.02)
 (-3.85,0.01)
 (-3.84,0.01)
 (-3.83,0)
 (-3.82,0)
 (-3.81,-0.01)
 (-3.8,-0.02)
 (-3.79,-0.02)
 (-3.78,-0.03)
 (-3.77,-0.04)
 (-3.76,-0.04)
 (-3.75,-0.05)
 (-3.74,-0.06)
 (-3.73,-0.06)
 (-3.72,-0.07)
 (-3.71,-0.08)
 (-3.7,-0.08)
 (-3.69,-0.09)
 (-3.68,-0.1)
 (-3.67,-0.1)
 (-3.66,-0.11)
 (-3.65,-0.12)
 (-3.64,-0.12)
 (-3.63,-0.13)
 (-3.62,-0.14)
 (-3.61,-0.14)
 (-3.6,-0.15)
 (-3.59,-0.16)
 (-3.58,-0.16)
 (-3.57,-0.17)
 (-3.56,-0.18)
 (-3.55,-0.18)
 (-3.54,-0.19)
 (-3.53,-0.2)
 (-3.52,-0.2)
 (-3.51,-0.21)
 (-3.5,-0.22)
 (-3.49,-0.22)
 (-3.48,-0.23)
 (-3.47,-0.24)
 (-3.46,-0.24)
 (-3.45,-0.25)
 (-3.44,-0.26)
 (-3.43,-0.26)
 (-3.42,-0.27)
 (-3.41,-0.28)
 (-3.4,-0.29)
 (-3.39,-0.29)
 (-3.38,-0.3)
 (-3.37,-0.31)
 (-3.36,-0.31)
 (-3.35,-0.32)
 (-3.34,-0.33)
 (-3.33,-0.33)
 (-3.32,-0.34)
 (-3.31,-0.35)
 (-3.3,-0.36)
 (-3.29,-0.36)
 (-3.28,-0.37)
 (-3.27,-0.38)
 (-3.26,-0.38)
 (-3.25,-0.39)
 (-3.24,-0.4)
 (-3.23,-0.41)
 (-3.22,-0.41)
 (-3.21,-0.42)
 (-3.2,-0.43)
 (-3.19,-0.44)
 (-3.18,-0.44)
 (-3.17,-0.45)
 (-3.16,-0.46)
 (-3.15,-0.46)
 (-3.14,-0.47)
 (-3.13,-0.48)
 (-3.12,-0.49)
 (-3.11,-0.49)
 (-3.1,-0.5)
 (-3.09,-0.51)
 (-3.08,-0.52)
 (-3.07,-0.52)
 (-3.06,-0.53)
 (-3.05,-0.54)
 (-3.04,-0.55)
 (-3.03,-0.55)
 (-3.02,-0.56)
 (-3.01,-0.57)
 (-3,-0.58)
 (-2.99,-0.58)
 (-2.98,-0.59)
 (-2.97,-0.6)
 (-2.96,-0.61)
 (-2.95,-0.61)
 (-2.94,-0.62)
 (-2.93,-0.63)
 (-2.92,-0.64)
 (-2.91,-0.64)
 (-2.9,-0.65)
 (-2.89,-0.66)
 (-2.88,-0.67)
 (-2.87,-0.68)
 (-2.86,-0.68)
 (-2.85,-0.69)
 (-2.84,-0.7)
 (-2.83,-0.71)
 (-2.82,-0.72)
 (-2.81,-0.72)
 (-2.8,-0.73)
 (-2.79,-0.74)
 (-2.78,-0.75)
 (-2.77,-0.75)
 (-2.76,-0.76)
 (-2.75,-0.77)
 (-2.74,-0.78)
 (-2.73,-0.79)
 (-2.72,-0.79)
 (-2.71,-0.8)
 (-2.7,-0.81)
 (-2.69,-0.82)
 (-2.68,-0.83)
 (-2.67,-0.83)
 (-2.66,-0.84)
 (-2.65,-0.85)
 (-2.64,-0.86)
 (-2.63,-0.87)
 (-2.62,-0.88)
 (-2.61,-0.88)
 (-2.6,-0.89)
 (-2.59,-0.9)
 (-2.58,-0.91)
 (-2.57,-0.92)
 (-2.56,-0.92)
 (-2.55,-0.93)
 (-2.54,-0.94)
 (-2.53,-0.95)
 (-2.52,-0.96)
 (-2.51,-0.97)
 (-2.5,-0.98)
 (-2.49,-0.98)
 (-2.48,-0.99)
 (-2.47,-1)
 (-2.46,-1.01)
 (-2.45,-1.02)
 (-2.44,-1.03)
 (-2.43,-1.03)
 (-2.42,-1.04)
 (-2.41,-1.05)
 (-2.4,-1.06)
 (-2.39,-1.07)
 (-2.38,-1.08)
 (-2.37,-1.09)
 (-2.36,-1.09)
 (-2.35,-1.1)
 (-2.34,-1.11)
 (-2.33,-1.12)
 (-2.32,-1.13)
 (-2.31,-1.14)
 (-2.3,-1.15)
 (-2.29,-1.16)
 (-2.28,-1.16)
 (-2.27,-1.17)
 (-2.26,-1.18)
 (-2.25,-1.19)
 (-2.24,-1.2)
 (-2.23,-1.21)
 (-2.22,-1.22)
 (-2.21,-1.23)
 (-2.2,-1.24)
 (-2.19,-1.24)
 (-2.18,-1.25)
 (-2.17,-1.26)
 (-2.16,-1.27)
 (-2.15,-1.28)
 (-2.14,-1.29)
 (-2.13,-1.3)
 (-2.12,-1.31)
 (-2.11,-1.32)
 (-2.1,-1.33)
 (-2.09,-1.34)
 (-2.08,-1.34)
 (-2.07,-1.35)
 (-2.06,-1.36)
 (-2.05,-1.37)
 (-2.04,-1.38)
 (-2.03,-1.39)
 (-2.02,-1.4)
 (-2.01,-1.41)
 (-2,-1.42)
 (-1.99,-1.43)
 (-1.98,-1.44)
 (-1.97,-1.45)
 (-1.96,-1.46)
 (-1.95,-1.47)
 (-1.94,-1.48)
 (-1.93,-1.48)
 (-1.92,-1.49)
 (-1.91,-1.5)
 (-1.9,-1.51)
 (-1.89,-1.52)
 (-1.88,-1.53)
 (-1.87,-1.54)
 (-1.86,-1.55)
 (-1.85,-1.56)
 (-1.84,-1.57)
 (-1.83,-1.58)
 (-1.82,-1.59)
 (-1.81,-1.6)
 (-1.8,-1.61)
 (-1.79,-1.62)
 (-1.78,-1.63)
 (-1.77,-1.64)
 (-1.76,-1.65)
 (-1.75,-1.66)
 (-1.74,-1.67)
 (-1.73,-1.68)
 (-1.72,-1.69)
 (-1.71,-1.7)
 (-1.7,-1.71)
 (-1.69,-1.72)
 (-1.68,-1.73)
 (-1.67,-1.74)
 (-1.66,-1.75)
 (-1.65,-1.76)
 (-1.64,-1.77)
 (-1.63,-1.78)
 (-1.62,-1.79)
 (-1.61,-1.8)
 (-1.6,-1.81)
 (-1.59,-1.82)
 (-1.58,-1.83)
 (-1.57,-1.84)
 (-1.56,-1.85)
 (-1.55,-1.86)
 (-1.54,-1.87)
 (-1.53,-1.88)
 (-1.52,-1.89)
 (-1.51,-1.9)
 (-1.5,-1.91)
 (-1.49,-1.92)
 (-1.48,-1.93)
 (-1.48,-1.94)
 (-1.47,-1.95)
 (-1.46,-1.96)
 (-1.45,-1.97)
 (-1.44,-1.98)
 (-1.43,-1.99)
 (-1.42,-2)
 (-1.41,-2.01)
 (-1.4,-2.02)
 (-1.39,-2.03)
 (-1.38,-2.04)
 (-1.37,-2.05)
 (-1.36,-2.06)
 (-1.35,-2.07)
 (-1.34,-2.08)
 (-1.34,-2.09)
 (-1.33,-2.1)
 (-1.32,-2.11)
 (-1.31,-2.12)
 (-1.3,-2.13)
 (-1.29,-2.14)
 (-1.28,-2.15)
 (-1.27,-2.16)
 (-1.26,-2.17)
 (-1.25,-2.18)
 (-1.24,-2.19)
 (-1.24,-2.2)
 (-1.23,-2.21)
 (-1.22,-2.22)
 (-1.21,-2.23)
 (-1.2,-2.24)
 (-1.19,-2.25)
 (-1.18,-2.26)
 (-1.17,-2.27)
 (-1.16,-2.28)
 (-1.16,-2.29)
 (-1.15,-2.3)
 (-1.14,-2.31)
 (-1.13,-2.32)
 (-1.12,-2.33)
 (-1.11,-2.34)
 (-1.1,-2.35)
 (-1.09,-2.36)
 (-1.09,-2.37)
 (-1.08,-2.38)
 (-1.07,-2.39)
 (-1.06,-2.4)
 (-1.05,-2.41)
 (-1.04,-2.42)
 (-1.03,-2.43)
 (-1.03,-2.44)
 (-1.02,-2.45)
 (-1.01,-2.46)
 (-1,-2.47)
 (-0.99,-2.48)
 (-0.98,-2.49)
 (-0.98,-2.5)
 (-0.97,-2.51)
 (-0.96,-2.52)
 (-0.95,-2.53)
 (-0.94,-2.54)
 (-0.93,-2.55)
 (-0.92,-2.56)
 (-0.92,-2.57)
 (-0.91,-2.58)
 (-0.9,-2.59)
 (-0.89,-2.6)
 (-0.88,-2.61)
 (-0.88,-2.62)
 (-0.87,-2.63)
 (-0.86,-2.64)
 (-0.85,-2.65)
 (-0.84,-2.66)
 (-0.83,-2.67)
 (-0.83,-2.68)
 (-0.82,-2.69)
 (-0.81,-2.7)
 (-0.8,-2.71)
 (-0.79,-2.72)
 (-0.79,-2.73)
 (-0.78,-2.74)
 (-0.77,-2.75)
 (-0.76,-2.76)
 (-0.75,-2.77)
 (-0.75,-2.78)
 (-0.74,-2.79)
 (-0.73,-2.8)
 (-0.72,-2.81)
 (-0.72,-2.82)
 (-0.71,-2.83)
 (-0.7,-2.84)
 (-0.69,-2.85)
 (-0.68,-2.86)
 (-0.68,-2.87)
 (-0.67,-2.88)
 (-0.66,-2.89)
 (-0.65,-2.9)
 (-0.64,-2.91)
 (-0.64,-2.92)
 (-0.63,-2.93)
 (-0.62,-2.94)
 (-0.61,-2.95)
 (-0.61,-2.96)
 (-0.6,-2.97)
 (-0.59,-2.98)
 (-0.58,-2.99)
 (-0.58,-3)
 (-0.57,-3.01)
 (-0.56,-3.02)
 (-0.55,-3.03)
 (-0.55,-3.04)
 (-0.54,-3.05)
 (-0.53,-3.06)
 (-0.52,-3.07)
 (-0.52,-3.08)
 (-0.51,-3.09)
 (-0.5,-3.1)
 (-0.49,-3.11)
 (-0.49,-3.12)
 (-0.48,-3.13)
 (-0.47,-3.14)
 (-0.46,-3.15)
 (-0.46,-3.16)
 (-0.45,-3.17)
 (-0.44,-3.18)
 (-0.44,-3.19)
 (-0.43,-3.2)
 (-0.42,-3.21)
 (-0.41,-3.22)
 (-0.41,-3.23)
 (-0.4,-3.24)
 (-0.39,-3.25)
 (-0.38,-3.26)
 (-0.38,-3.27)
 (-0.37,-3.28)
 (-0.36,-3.29)
 (-0.36,-3.3)
 (-0.35,-3.31)
 (-0.34,-3.32)
 (-0.33,-3.33)
 (-0.33,-3.34)
 (-0.32,-3.35)
 (-0.31,-3.36)
 (-0.31,-3.37)
 (-0.3,-3.38)
 (-0.29,-3.39)
 (-0.29,-3.4)
 (-0.28,-3.41)
 (-0.27,-3.42)
 (-0.26,-3.43)
 (-0.26,-3.44)
 (-0.25,-3.45)
 (-0.24,-3.46)
 (-0.24,-3.47)
 (-0.23,-3.48)
 (-0.22,-3.49)
 (-0.22,-3.5)
 (-0.21,-3.51)
 (-0.2,-3.52)
 (-0.2,-3.53)
 (-0.19,-3.54)
 (-0.18,-3.55)
 (-0.18,-3.56)
 (-0.17,-3.57)
 (-0.16,-3.58)
 (-0.16,-3.59)
 (-0.15,-3.6)
 (-0.14,-3.61)
 (-0.14,-3.62)
 (-0.13,-3.63)
 (-0.12,-3.64)
 (-0.12,-3.65)
 (-0.11,-3.66)
 (-0.1,-3.67)
 (-0.1,-3.68)
 (-0.09,-3.69)
 (-0.08,-3.7)
 (-0.08,-3.71)
 (-0.07,-3.72)
 (-0.06,-3.73)
 (-0.06,-3.74)
 (-0.05,-3.75)
 (-0.04,-3.76)
 (-0.04,-3.77)
 (-0.03,-3.78)
 (-0.02,-3.79)
 (-0.02,-3.8)
 (-0.01,-3.81)
 (0,-3.82)
 (0,-3.83)
 (0.01,-3.84)
 (0.01,-3.85)
 (0.02,-3.86)
 (0.03,-3.87)
 (0.03,-3.88)
 (0.04,-3.89)
 (0.05,-3.9)
 (0.05,-3.91)
 (0.06,-3.92)
 (0.07,-3.93)
 (0.07,-3.94)
 (0.08,-3.95)
 (0.08,-3.96)
 (0.09,-3.97)
 (0.1,-3.98)
 (0.1,-3.99)
 (0.11,-4)
 (0.12,-4.01)
 (0.12,-4.02)
 (0.13,-4.03)
 (0.13,-4.04)
 (0.14,-4.05)
 (0.15,-4.06)
 (0.15,-4.07)
 (0.16,-4.08)
 (0.16,-4.09)
 (0.17,-4.1)
 (0.18,-4.11)
 (0.18,-4.12)
 (0.19,-4.13)
 (0.2,-4.13)
 (3,-4.13)

};
\addlegendentry{\Large SW Region for rate $0.5$}

\addplot [
line width=0.2pt,fill=gray,fill opacity=0.2
]
coordinates{
 (-4.13,3)
 (-4.13,0.2)
 (-4.12,0.19)
 (-4.11,0.19)
 (-4.1,0.18)
 (-4.09,0.17)
 (-4.08,0.17)
 (-4.07,0.16)
 (-4.06,0.16)
 (-4.05,0.15)
 (-4.04,0.14)
 (-4.03,0.14)
 (-4.02,0.13)
 (-4.01,0.13)
 (-4,0.12)
 (-3.99,0.11)
 (-3.98,0.11)
 (-3.97,0.1)
 (-3.96,0.09)
 (-3.95,0.09)
 (-3.94,0.08)
 (-3.93,0.08)
 (-3.92,0.07)
 (-3.91,0.06)
 (-3.9,0.06)
 (-3.89,0.05)
 (-3.88,0.04)
 (-3.87,0.04)
 (-3.86,0.03)
 (-3.85,0.02)
 (-3.84,0.02)
 (-3.83,0.01)
 (-3.82,0.01)
 (-3.81,0)
 (-3.8,-0.01)
 (-3.79,-0.01)
 (-3.78,-0.02)
 (-3.77,-0.03)
 (-3.76,-0.03)
 (-3.75,-0.04)
 (-3.74,-0.05)
 (-3.73,-0.05)
 (-3.72,-0.06)
 (-3.71,-0.07)
 (-3.7,-0.07)
 (-3.69,-0.08)
 (-3.68,-0.09)
 (-3.67,-0.09)
 (-3.66,-0.1)
 (-3.65,-0.11)
 (-3.64,-0.11)
 (-3.63,-0.12)
 (-3.62,-0.13)
 (-3.61,-0.13)
 (-3.6,-0.14)
 (-3.59,-0.15)
 (-3.58,-0.15)
 (-3.57,-0.16)
 (-3.56,-0.17)
 (-3.55,-0.17)
 (-3.54,-0.18)
 (-3.53,-0.19)
 (-3.52,-0.19)
 (-3.51,-0.2)
 (-3.5,-0.21)
 (-3.49,-0.21)
 (-3.48,-0.22)
 (-3.47,-0.23)
 (-3.46,-0.23)
 (-3.45,-0.24)
 (-3.44,-0.25)
 (-3.43,-0.25)
 (-3.42,-0.26)
 (-3.41,-0.27)
 (-3.4,-0.28)
 (-3.39,-0.28)
 (-3.38,-0.29)
 (-3.37,-0.3)
 (-3.36,-0.3)
 (-3.35,-0.31)
 (-3.34,-0.32)
 (-3.33,-0.32)
 (-3.32,-0.33)
 (-3.31,-0.34)
 (-3.3,-0.35)
 (-3.29,-0.35)
 (-3.28,-0.36)
 (-3.27,-0.37)
 (-3.26,-0.37)
 (-3.25,-0.38)
 (-3.24,-0.39)
 (-3.23,-0.4)
 (-3.22,-0.4)
 (-3.21,-0.41)
 (-3.2,-0.42)
 (-3.19,-0.43)
 (-3.18,-0.43)
 (-3.17,-0.44)
 (-3.16,-0.45)
 (-3.15,-0.45)
 (-3.14,-0.46)
 (-3.13,-0.47)
 (-3.12,-0.48)
 (-3.11,-0.48)
 (-3.1,-0.49)
 (-3.09,-0.5)
 (-3.08,-0.51)
 (-3.07,-0.51)
 (-3.06,-0.52)
 (-3.05,-0.53)
 (-3.04,-0.54)
 (-3.03,-0.54)
 (-3.02,-0.55)
 (-3.01,-0.56)
 (-3,-0.57)
 (-2.99,-0.57)
 (-2.98,-0.58)
 (-2.97,-0.59)
 (-2.96,-0.6)
 (-2.95,-0.6)
 (-2.94,-0.61)
 (-2.93,-0.62)
 (-2.92,-0.63)
 (-2.91,-0.63)
 (-2.9,-0.64)
 (-2.89,-0.65)
 (-2.88,-0.66)
 (-2.87,-0.67)
 (-2.86,-0.67)
 (-2.85,-0.68)
 (-2.84,-0.69)
 (-2.83,-0.7)
 (-2.82,-0.7)
 (-2.81,-0.71)
 (-2.8,-0.72)
 (-2.79,-0.73)
 (-2.78,-0.74)
 (-2.77,-0.74)
 (-2.76,-0.75)
 (-2.75,-0.76)
 (-2.74,-0.77)
 (-2.73,-0.78)
 (-2.72,-0.78)
 (-2.71,-0.79)
 (-2.7,-0.8)
 (-2.69,-0.81)
 (-2.68,-0.82)
 (-2.67,-0.82)
 (-2.66,-0.83)
 (-2.65,-0.84)
 (-2.64,-0.85)
 (-2.63,-0.86)
 (-2.62,-0.86)
 (-2.61,-0.87)
 (-2.6,-0.88)
 (-2.59,-0.89)
 (-2.58,-0.9)
 (-2.57,-0.91)
 (-2.56,-0.91)
 (-2.55,-0.92)
 (-2.54,-0.93)
 (-2.53,-0.94)
 (-2.52,-0.95)
 (-2.51,-0.96)
 (-2.5,-0.96)
 (-2.49,-0.97)
 (-2.48,-0.98)
 (-2.47,-0.99)
 (-2.46,-1)
 (-2.45,-1.01)
 (-2.44,-1.02)
 (-2.43,-1.02)
 (-2.42,-1.03)
 (-2.41,-1.04)
 (-2.4,-1.05)
 (-2.39,-1.06)
 (-2.38,-1.07)
 (-2.37,-1.07)
 (-2.36,-1.08)
 (-2.35,-1.09)
 (-2.34,-1.1)
 (-2.33,-1.11)
 (-2.32,-1.12)
 (-2.31,-1.13)
 (-2.3,-1.14)
 (-2.29,-1.14)
 (-2.28,-1.15)
 (-2.27,-1.16)
 (-2.26,-1.17)
 (-2.25,-1.18)
 (-2.24,-1.19)
 (-2.23,-1.2)
 (-2.22,-1.21)
 (-2.21,-1.22)
 (-2.2,-1.22)
 (-2.19,-1.23)
 (-2.18,-1.24)
 (-2.17,-1.25)
 (-2.16,-1.26)
 (-2.15,-1.27)
 (-2.14,-1.28)
 (-2.13,-1.29)
 (-2.12,-1.3)
 (-2.11,-1.31)
 (-2.1,-1.31)
 (-2.09,-1.32)
 (-2.08,-1.33)
 (-2.07,-1.34)
 (-2.06,-1.35)
 (-2.05,-1.36)
 (-2.04,-1.37)
 (-2.03,-1.38)
 (-2.02,-1.39)
 (-2.01,-1.4)
 (-2,-1.41)
 (-1.99,-1.42)
 (-1.98,-1.43)
 (-1.97,-1.44)
 (-1.96,-1.44)
 (-1.95,-1.45)
 (-1.94,-1.46)
 (-1.93,-1.47)
 (-1.92,-1.48)
 (-1.91,-1.49)
 (-1.9,-1.5)
 (-1.89,-1.51)
 (-1.88,-1.52)
 (-1.87,-1.53)
 (-1.86,-1.54)
 (-1.85,-1.55)
 (-1.84,-1.56)
 (-1.83,-1.57)
 (-1.82,-1.58)
 (-1.81,-1.59)
 (-1.8,-1.6)
 (-1.79,-1.61)
 (-1.78,-1.62)
 (-1.77,-1.63)
 (-1.76,-1.64)
 (-1.75,-1.65)
 (-1.74,-1.66)
 (-1.73,-1.67)
 (-1.72,-1.68)
 (-1.71,-1.69)
 (-1.7,-1.7)
 (-1.69,-1.71)
 (-1.68,-1.72)
 (-1.67,-1.73)
 (-1.66,-1.74)
 (-1.65,-1.75)
 (-1.64,-1.76)
 (-1.63,-1.77)
 (-1.62,-1.78)
 (-1.61,-1.79)
 (-1.6,-1.8)
 (-1.59,-1.81)
 (-1.58,-1.82)
 (-1.57,-1.83)
 (-1.56,-1.84)
 (-1.55,-1.85)
 (-1.54,-1.86)
 (-1.53,-1.87)
 (-1.52,-1.88)
 (-1.51,-1.89)
 (-1.5,-1.9)
 (-1.49,-1.91)
 (-1.48,-1.92)
 (-1.47,-1.93)
 (-1.46,-1.94)
 (-1.45,-1.95)
 (-1.44,-1.96)
 (-1.44,-1.97)
 (-1.43,-1.98)
 (-1.42,-1.99)
 (-1.41,-2)
 (-1.4,-2.01)
 (-1.39,-2.02)
 (-1.38,-2.03)
 (-1.37,-2.04)
 (-1.36,-2.05)
 (-1.35,-2.06)
 (-1.34,-2.07)
 (-1.33,-2.08)
 (-1.32,-2.09)
 (-1.31,-2.1)
 (-1.31,-2.11)
 (-1.3,-2.12)
 (-1.29,-2.13)
 (-1.28,-2.14)
 (-1.27,-2.15)
 (-1.26,-2.16)
 (-1.25,-2.17)
 (-1.24,-2.18)
 (-1.23,-2.19)
 (-1.22,-2.2)
 (-1.22,-2.21)
 (-1.21,-2.22)
 (-1.2,-2.23)
 (-1.19,-2.24)
 (-1.18,-2.25)
 (-1.17,-2.26)
 (-1.16,-2.27)
 (-1.15,-2.28)
 (-1.14,-2.29)
 (-1.14,-2.3)
 (-1.13,-2.31)
 (-1.12,-2.32)
 (-1.11,-2.33)
 (-1.1,-2.34)
 (-1.09,-2.35)
 (-1.08,-2.36)
 (-1.07,-2.37)
 (-1.07,-2.38)
 (-1.06,-2.39)
 (-1.05,-2.4)
 (-1.04,-2.41)
 (-1.03,-2.42)
 (-1.02,-2.43)
 (-1.02,-2.44)
 (-1.01,-2.45)
 (-1,-2.46)
 (-0.99,-2.47)
 (-0.98,-2.48)
 (-0.97,-2.49)
 (-0.96,-2.5)
 (-0.96,-2.51)
 (-0.95,-2.52)
 (-0.94,-2.53)
 (-0.93,-2.54)
 (-0.92,-2.55)
 (-0.91,-2.56)
 (-0.91,-2.57)
 (-0.9,-2.58)
 (-0.89,-2.59)
 (-0.88,-2.6)
 (-0.87,-2.61)
 (-0.86,-2.62)
 (-0.86,-2.63)
 (-0.85,-2.64)
 (-0.84,-2.65)
 (-0.83,-2.66)
 (-0.82,-2.67)
 (-0.82,-2.68)
 (-0.81,-2.69)
 (-0.8,-2.7)
 (-0.79,-2.71)
 (-0.78,-2.72)
 (-0.78,-2.73)
 (-0.77,-2.74)
 (-0.76,-2.75)
 (-0.75,-2.76)
 (-0.74,-2.77)
 (-0.74,-2.78)
 (-0.73,-2.79)
 (-0.72,-2.8)
 (-0.71,-2.81)
 (-0.7,-2.82)
 (-0.7,-2.83)
 (-0.69,-2.84)
 (-0.68,-2.85)
 (-0.67,-2.86)
 (-0.67,-2.87)
 (-0.66,-2.88)
 (-0.65,-2.89)
 (-0.64,-2.9)
 (-0.63,-2.91)
 (-0.63,-2.92)
 (-0.62,-2.93)
 (-0.61,-2.94)
 (-0.6,-2.95)
 (-0.6,-2.96)
 (-0.59,-2.97)
 (-0.58,-2.98)
 (-0.57,-2.99)
 (-0.57,-3)
 (-0.56,-3.01)
 (-0.55,-3.02)
 (-0.54,-3.03)
 (-0.54,-3.04)
 (-0.53,-3.05)
 (-0.52,-3.06)
 (-0.51,-3.07)
 (-0.51,-3.08)
 (-0.5,-3.09)
 (-0.49,-3.1)
 (-0.48,-3.11)
 (-0.48,-3.12)
 (-0.47,-3.13)
 (-0.46,-3.14)
 (-0.45,-3.15)
 (-0.45,-3.16)
 (-0.44,-3.17)
 (-0.43,-3.18)
 (-0.43,-3.19)
 (-0.42,-3.2)
 (-0.41,-3.21)
 (-0.4,-3.22)
 (-0.4,-3.23)
 (-0.39,-3.24)
 (-0.38,-3.25)
 (-0.37,-3.26)
 (-0.37,-3.27)
 (-0.36,-3.28)
 (-0.35,-3.29)
 (-0.35,-3.3)
 (-0.34,-3.31)
 (-0.33,-3.32)
 (-0.32,-3.33)
 (-0.32,-3.34)
 (-0.31,-3.35)
 (-0.3,-3.36)
 (-0.3,-3.37)
 (-0.29,-3.38)
 (-0.28,-3.39)
 (-0.28,-3.4)
 (-0.27,-3.41)
 (-0.26,-3.42)
 (-0.25,-3.43)
 (-0.25,-3.44)
 (-0.24,-3.45)
 (-0.23,-3.46)
 (-0.23,-3.47)
 (-0.22,-3.48)
 (-0.21,-3.49)
 (-0.21,-3.5)
 (-0.2,-3.51)
 (-0.19,-3.52)
 (-0.19,-3.53)
 (-0.18,-3.54)
 (-0.17,-3.55)
 (-0.17,-3.56)
 (-0.16,-3.57)
 (-0.15,-3.58)
 (-0.15,-3.59)
 (-0.14,-3.6)
 (-0.13,-3.61)
 (-0.13,-3.62)
 (-0.12,-3.63)
 (-0.11,-3.64)
 (-0.11,-3.65)
 (-0.1,-3.66)
 (-0.09,-3.67)
 (-0.09,-3.68)
 (-0.08,-3.69)
 (-0.07,-3.7)
 (-0.07,-3.71)
 (-0.06,-3.72)
 (-0.05,-3.73)
 (-0.05,-3.74)
 (-0.04,-3.75)
 (-0.03,-3.76)
 (-0.03,-3.77)
 (-0.02,-3.78)
 (-0.01,-3.79)
 (-0.01,-3.8)
 (0,-3.81)
 (0.01,-3.82)
 (0.01,-3.83)
 (0.02,-3.84)
 (0.02,-3.85)
 (0.03,-3.86)
 (0.04,-3.87)
 (0.04,-3.88)
 (0.05,-3.89)
 (0.06,-3.9)
 (0.06,-3.91)
 (0.07,-3.92)
 (0.08,-3.93)
 (0.08,-3.94)
 (0.09,-3.95)
 (0.09,-3.96)
 (0.1,-3.97)
 (0.11,-3.98)
 (0.11,-3.99)
 (0.12,-4)
 (0.13,-4.01)
 (0.13,-4.02)
 (0.14,-4.03)
 (0.14,-4.04)
 (0.15,-4.05)
 (0.16,-4.06)
 (0.16,-4.07)
 (0.17,-4.08)
 (0.17,-4.09)
 (0.18,-4.1)
 (0.19,-4.11)
 (0.19,-4.12)
 (0.2,-4.13)
 (0.21,-4.13)
 (0.22,-4.13)
 (0.23,-4.13)
 (0.24,-4.13)
 (0.25,-4.13)
 (0.26,-4.13)
 (0.27,-4.13)
 (0.28,-4.13)
 (0.29,-4.13)
 (0.3,-4.13)
 (0.31,-4.13)
 (0.32,-4.13)
 (0.33,-4.13)
 (0.34,-4.13)
 (0.35,-4.13)
 (0.36,-4.13)
 (0.37,-4.13)
 (0.38,-4.13)
 (0.39,-4.13)
 (0.4,-4.13)
 (0.41,-4.13)
 (0.42,-4.13)
 (0.43,-4.13)
 (0.44,-4.13)
 (0.45,-4.13)
 (0.46,-4.13)
 (0.47,-4.13)
 (0.48,-4.13)
 (0.49,-4.13)
 (0.5,-4.13)
 (0.51,-4.13)
 (0.52,-4.13)
 (0.53,-4.13)
 (0.54,-4.13)
 (0.55,-4.13)
 (0.56,-4.13)
 (0.57,-4.13)
 (0.58,-4.13)
 (0.59,-4.13)
 (0.6,-4.13)
 (0.61,-4.13)
 (0.62,-4.13)
 (0.63,-4.13)
 (0.64,-4.13)
 (0.65,-4.13)
 (0.66,-4.13)
 (0.67,-4.13)
 (0.68,-4.13)
 (0.69,-4.13)
 (0.7,-4.13)
 (0.71,-4.13)
 (0.72,-4.13)
 (0.73,-4.13)
 (0.74,-4.13)
 (0.75,-4.13)
 (0.76,-4.13)
 (0.77,-4.13)
 (0.78,-4.13)
 (0.79,-4.13)
 (0.8,-4.13)
 (0.81,-4.13)
 (0.82,-4.13)
 (0.83,-4.13)
 (0.84,-4.13)
 (0.85,-4.13)
 (0.86,-4.13)
 (0.87,-4.13)
 (0.88,-4.13)
 (0.89,-4.13)
 (0.9,-4.13)
 (0.91,-4.13)
 (0.92,-4.13)
 (0.93,-4.13)
 (0.94,-4.13)
 (0.95,-4.13)
 (0.96,-4.13)
 (0.97,-4.13)
 (0.98,-4.13)
 (0.99,-4.13)
 (1,-4.13)
 (1.01,-4.13)
 (1.02,-4.13)
 (1.03,-4.13)
 (1.04,-4.13)
 (1.05,-4.13)
 (1.06,-4.13)
 (1.07,-4.13)
 (1.08,-4.13)
 (1.09,-4.13)
 (1.1,-4.13)
 (1.11,-4.13)
 (1.12,-4.13)
 (1.13,-4.13)
 (1.14,-4.13)
 (1.15,-4.13)
 (1.16,-4.13)
 (1.17,-4.13)
 (1.18,-4.13)
 (1.19,-4.13)
 (1.2,-4.13)
 (1.21,-4.13)
 (1.22,-4.13)
 (1.23,-4.13)
 (1.24,-4.13)
 (1.25,-4.13)
 (1.26,-4.13)
 (1.27,-4.13)
 (1.28,-4.13)
 (1.29,-4.13)
 (1.3,-4.13)
 (1.31,-4.13)
 (1.32,-4.13)
 (1.33,-4.13)
 (1.34,-4.13)
 (1.35,-4.13)
 (1.36,-4.13)
 (1.37,-4.13)
 (1.38,-4.13)
 (1.39,-4.13)
 (1.4,-4.13)
 (1.41,-4.13)
 (1.42,-4.13)
 (1.43,-4.13)
 (1.44,-4.13)
 (1.45,-4.13)
 (1.46,-4.13)
 (1.47,-4.13)
 (1.48,-4.13)
 (1.49,-4.13)
 (1.5,-4.13)
 (1.51,-4.13)
 (1.52,-4.13)
 (1.53,-4.13)
 (1.54,-4.13)
 (1.55,-4.13)
 (1.56,-4.13)
 (1.57,-4.13)
 (1.58,-4.13)
 (1.59,-4.13)
 (1.6,-4.13)
 (1.61,-4.13)
 (1.62,-4.13)
 (1.63,-4.13)
 (1.64,-4.13)
 (1.65,-4.13)
 (1.66,-4.13)
 (1.67,-4.13)
 (1.68,-4.13)
 (1.69,-4.13)
 (1.7,-4.13)
 (1.71,-4.13)
 (1.72,-4.13)
 (1.73,-4.13)
 (1.74,-4.13)
 (1.75,-4.13)
 (1.76,-4.13)
 (1.77,-4.13)
 (1.78,-4.13)
 (1.79,-4.13)
 (1.8,-4.13)
 (1.81,-4.13)
 (1.82,-4.13)
 (1.83,-4.13)
 (1.84,-4.13)
 (1.85,-4.13)
 (1.86,-4.13)
 (1.87,-4.13)
 (1.88,-4.13)
 (1.89,-4.13)
 (1.9,-4.13)
 (1.91,-4.13)
 (1.92,-4.13)
 (1.93,-4.13)
 (1.94,-4.13)
 (1.95,-4.13)
 (1.96,-4.13)
 (1.97,-4.13)
 (1.98,-4.13)
 (1.99,-4.13)
 (2,-4.13)
 (2.01,-4.13)
 (2.02,-4.13)
 (2.03,-4.13)
 (2.04,-4.13)
 (2.05,-4.13)
 (2.06,-4.13)
 (2.07,-4.13)
 (2.08,-4.13)
 (2.09,-4.13)
 (2.1,-4.13)
 (2.11,-4.13)
 (2.12,-4.13)
 (2.13,-4.13)
 (2.14,-4.13)
 (2.15,-4.13)
 (2.16,-4.13)
 (2.17,-4.13)
 (2.18,-4.13)
 (2.19,-4.13)
 (2.2,-4.13)
 (2.21,-4.13)
 (2.22,-4.13)
 (2.23,-4.13)
 (2.24,-4.13)
 (2.25,-4.13)
 (2.26,-4.13)
 (2.27,-4.13)
 (2.28,-4.13)
 (2.29,-4.13)
 (2.3,-4.13)
 (2.31,-4.13)
 (2.32,-4.13)
 (2.33,-4.13)
 (2.34,-4.13)
 (2.35,-4.13)
 (2.36,-4.13)
 (2.37,-4.13)
 (2.38,-4.13)
 (2.39,-4.13)
 (2.4,-4.13)
 (2.41,-4.13)
 (2.42,-4.13)
 (2.43,-4.13)
 (2.44,-4.13)
 (2.45,-4.13)
 (2.46,-4.13)
 (2.47,-4.13)
 (2.48,-4.13)
 (2.49,-4.13)
 (2.5,-4.13)
 (2.51,-4.13)
 (2.52,-4.13)
 (2.53,-4.13)
 (2.54,-4.13)
 (2.55,-4.13)
 (2.56,-4.13)
 (2.57,-4.13)
 (2.58,-4.13)
 (2.59,-4.13)
 (2.6,-4.13)
 (2.61,-4.13)
 (2.62,-4.13)
 (2.63,-4.13)
 (2.64,-4.13)
 (2.65,-4.13)
 (2.66,-4.13)
 (2.67,-4.13)
 (2.68,-4.13)
 (2.69,-4.13)
 (2.7,-4.13)
 (2.71,-4.13)
 (2.72,-4.13)
 (2.73,-4.13)
 (2.74,-4.13)
 (2.75,-4.13)
 (2.76,-4.13)
 (2.77,-4.13)
 (2.78,-4.13)
 (2.79,-4.13)
 (2.8,-4.13)
 (2.81,-4.13)
 (2.82,-4.13)
 (2.83,-4.13)
 (2.84,-4.13)
 (2.85,-4.13)
 (2.86,-4.13)
 (2.87,-4.13)
 (2.88,-4.13)
 (2.89,-4.13)
 (2.9,-4.13)
 (2.91,-4.13)
 (2.92,-4.13)
 (2.93,-4.13)
 (2.94,-4.13)
 (2.95,-4.13)
 (2.96,-4.13)
 (2.97,-4.13)
 (2.98,-4.13)
 (2.99,-4.13)
 (3,-4.13)

}|- (axis cs:3,3) -- cycle;
\addlegendentry{\Large ACPR (DE) - $(4,6,64,10)$}

\addplot [
black,fill=gray,fill opacity=0.4
]
coordinates{
 (-0.451172,3)
 (-0.451172,2.47852)
 (-0.399849,2.46852)
 (-0.338933,2.45852)
 (-0.275605,2.44852)
 (-0.219771,2.43852)
 (-0.164328,2.42852)
 (-0.109276,2.41852)
 (-0.0615953,2.40852)
 (-0.011474,2.39852)
 (0.0341263,2.38852)
 (0.0849213,2.37852)
 (0.127084,2.36852)
 (0.168935,2.35852)
 (0.210473,2.34852)
 (0.251699,2.33852)
 (0.292612,2.32852)
 (0.327127,2.31852)
 (0.365415,2.30852)
 (0.402735,2.29852)
 (0.435748,2.28852)
 (0.471164,2.27852)
 (0.502312,2.26852)
 (0.538499,2.25852)
 (0.566484,2.24852)
 (0.596861,2.23852)
 (0.626329,2.22852)
 (0.65946,2.21852)
 (0.686468,2.20852)
 (0.713242,2.19852)
 (0.739781,2.18852)
 (0.766086,2.17852)
 (0.792156,2.16852)
 (0.817993,2.15852)
 (0.843594,2.14852)
 (0.868962,2.13852)
 (0.894095,2.12852)
 (0.913347,2.11852)
 (0.938658,2.10852)
 (0.958128,2.09852)
 (0.987323,2.08852)
 (1.00142,2.07852)
 (1.02275,2.06852)
 (1.04876,2.05852)
 (1.06719,2.04852)
 (1.083,2.03852)
 (1.10832,2.02852)
 (1.12132,2.01852)
 (1.14138,2.00852)
 (1.166,1.99852)
 (1.17846,1.98852)
 (1.19551,1.97852)
 (1.21118,1.96852)
 (1.23373,1.95852)
 (1.24548,1.94852)
 (1.26175,1.93852)
 (1.27666,1.92852)
 (1.29832,1.91852)
 (1.30938,1.90852)
 (1.32486,1.89852)
 (1.33901,1.88852)
 (1.3598,1.87852)
 (1.37015,1.86852)
 (1.38485,1.85852)
 (1.39711,1.84852)
 (1.41814,1.83852)
 (1.42779,1.82852)
 (1.43728,1.81852)
 (1.45103,1.80852)
 (1.46238,1.79852)
 (1.48231,1.78852)
 (1.49109,1.77852)
 (1.49973,1.76852)
 (1.51252,1.75852)
 (1.52404,1.74852)
 (1.54179,1.73852)
 (1.54971,1.72852)
 (1.56172,1.71852)
 (1.56932,1.70852)
 (1.58043,1.69852)
 (1.59658,1.68852)
 (1.60364,1.67852)
 (1.6147,1.66852)
 (1.62349,1.65852)
 (1.64031,1.64852)
 (1.64852,1.64031)
 (1.65852,1.62349)
 (1.66852,1.6147)
 (1.67852,1.60364)
 (1.68852,1.59658)
 (1.69852,1.58043)
 (1.70852,1.56932)
 (1.71852,1.56172)
 (1.72852,1.54971)
 (1.73852,1.54179)
 (1.74852,1.52404)
 (1.75852,1.51252)
 (1.76852,1.49973)
 (1.77852,1.49109)
 (1.78852,1.48231)
 (1.79852,1.46238)
 (1.80852,1.45103)
 (1.81852,1.43728)
 (1.82852,1.42779)
 (1.83852,1.41814)
 (1.84852,1.39711)
 (1.85852,1.38485)
 (1.86852,1.37015)
 (1.87852,1.3598)
 (1.88852,1.33901)
 (1.89852,1.32486)
 (1.90852,1.30938)
 (1.91852,1.29832)
 (1.92852,1.27666)
 (1.93852,1.26175)
 (1.94852,1.24548)
 (1.95852,1.23373)
 (1.96852,1.21118)
 (1.97852,1.19551)
 (1.98852,1.17846)
 (1.99852,1.166)
 (2.00852,1.14138)
 (2.01852,1.12132)
 (2.02852,1.10832)
 (2.03852,1.083)
 (2.04852,1.06719)
 (2.05852,1.04876)
 (2.06852,1.02275)
 (2.07852,1.00142)
 (2.08852,0.987323)
 (2.09852,0.958128)
 (2.10852,0.938658)
 (2.11852,0.913347)
 (2.12852,0.894095)
 (2.13852,0.868962)
 (2.14852,0.843594)
 (2.15852,0.817993)
 (2.16852,0.792156)
 (2.17852,0.766086)
 (2.18852,0.739781)
 (2.19852,0.713242)
 (2.20852,0.686468)
 (2.21852,0.65946)
 (2.22852,0.626329)
 (2.23852,0.596861)
 (2.24852,0.566484)
 (2.25852,0.538499)
 (2.26852,0.502312)
 (2.27852,0.471164)
 (2.28852,0.435748)
 (2.29852,0.402735)
 (2.30852,0.365415)
 (2.31852,0.327127)
 (2.32852,0.292612)
 (2.33852,0.251699)
 (2.34852,0.210473)
 (2.35852,0.168935)
 (2.36852,0.127084)
 (2.37852,0.0849213)
 (2.38852,0.0341263)
 (2.39852,-0.011474)
 (2.40852,-0.0615953)
 (2.41852,-0.109276)
 (2.42852,-0.164328)
 (2.43852,-0.219771)
 (2.44852,-0.275605)
 (2.45852,-0.338933)
 (2.46852,-0.399849)
 (2.47852,-0.451172)
 (3,-0.451172)

}|- (axis cs:3,3) -- cycle;
\addlegendentry{\Large ACPR (DE) - (4,6)};

\end{axis}

\end{tikzpicture}

%% file: gaussian_mac.tex
Consider a $2$-user additive Gaussian multiple access channel (MAC),
given by
\begin{align*}
  Y = h_{1}X_{1} + h_{2}X_{2} + Z,
\end{align*}
with $Z\sim\mathcal{N}(0,1)$, $X_{1},X_{2}\in\{\pm 1\}$ and
$h_{1},h_{2}\in\mathbb{R}$. We assume that the fading coefficients are
not known at the transmitter and we consider the notion of
universality with respect to fading coefficients. The factor graph of
the joint decoder consists of two single user Tanner graphs, whose
variable nodes are connected through a function node
\cite[p. 308]{RU-2008}. 
Using the notation described in
Section~\ref{sec:dens-evol-exit}, the DE equation for the joint
decoder with symmetric fading coefficients is given by
\begin{align*}
  \a_{\ell+1} &=
  f\Bigl(L\left(\rho(\a_{\ell})\right),\a_{\text{BAWGNC}}\Bigr)\varoast
  \lambda(\rho(\a_{\ell})).
\end{align*}
Here, $f$ denotes the operation at the function node for
transformation of densities, and is chosen under the assumption of
transmission of a random coset of the LDPC code. Preliminary DE
results show that spatial coupling allows for near universal
performance on the Gaussian MAC (see
Fig.~\ref{fig:roc_ldpc_awgn_bac}). In our future work, we will derive
the area theorem (with an appropriate GEXIT kernel) for this problem
to formalize this result.


%% file: results.tex
It was shown in \cite{Kudekar-it11}, that for transmission over
erasure channels, the BP threshold of spatially-coupled ensembles is
essentially equal to the MAP threshold of the underlying
ensemble. This was observed numerically for general BMS channels in
\cite{Kudekar-istc10,Lentmaier-it05}. In this work, we numerically
show that the phenomenon of threshold saturation is very general and
can provide universality for multi-user scenarios. In particular, we
considered the noisy Slepian-Wolf problem and showed that spatial
coupling boosts the BP threshold of the joint decoder to the MAP
threshold of the underlying ensemble. The density evolution ACPRs for
the two scenarios considered in this paper are shown in
Fig.~\ref{fig:roc_ldpc_erasure} and \ref{fig:roc_ldpc_awgn}. These
figures show that spatially coupled ensembles are near universal for
this problem. The analytic proof of this result remains an open
problem. Such a proof would essentially show that it is possible to
achieve universality for the noisy Slepian-Wolf problem under
iterative decoding. 

